\documentclass[12pt,a4paper]{article}

\usepackage[english]{babel}
\usepackage{amsmath, amsfonts, amssymb, mathrsfs, dsfont}
\usepackage{titling}
\usepackage{graphicx}
\usepackage[caption = false]{subfig}
\usepackage{float}
\usepackage{authblk}
\usepackage[babel]{csquotes}
\usepackage{siunitx}

\usepackage[style=numeric-comp,sorting=none,url=false,maxbibnames=2, doi= false, eprint=false,useprefix,hyperref,backend=bibtex]{biblatex}
\usepackage{xcolor}
\usepackage{hyperref}

\newcommand{\ca}{\textit{ca.} } 

\newcommand{\fa}{\textit{(a)}} 
\newcommand{\fb}{\textit{(b)}} 
\newcommand{\fc}{\textit{(c)}} 
\newcommand{\fd}{\textit{(d)}} 

\newcommand{\CCd}{C$_2$(d$^3\Pi_g)$ } 
 
\newcommand{\Swan}{d$^3\Pi_g \rightarrow$ a$^3\Pi_u$}
\newcommand{\COB}{CO(B$^1\Sigma^+$) } 
\newcommand{\Angstrom}{B$^1\Sigma^+ \rightarrow$ A$^1\Pi$} 
\newcommand{\COO}{CO$_2$ }

\bibliography{DatabaseCitation.bib}

\title{Performance analysis of a 2.45~GHz microwave plasma torch for CO$_2$ decomposition in gas swirl configuration}
\author[]{F. A. D'Isa*}
\author[]{E. A. D. Carbone*}
\author[]{A. Hecimovic}
\author[]{U. Fantz}

\affil[]{ Max Planck Institute for Plasma Physics, \\Boltzmannstr. 2, 85748 Garching, Germany\\
{\small *Corresponding authors: federico.disa@ipp.mpg.de, \\ emile.carbone@ipp.mpg.de}}

\begin{document}

 
\maketitle

Submitted to: \textit{ChemSusChem}.

\begin{abstract}
Microwave plasmas are a promising technology for energy-efficient \COO valorization via conversion of \COO into CO and O$_2$ using renewable energies. A 2.45 GHz microwave plasma torch with swirling \COO gas flow is studied in a large pressure (60-1000~mbar) and flow (5-100~slm) range. Two different modes of the plasma torch, depending on the operating pressure and microwave input power, are described: at pressures below 120~mbar the plasma fills most of the plasma torch volume whereas at pressures of about 120~mbar an abrupt contraction of the plasma in the center of the resonator is observed along with an increase of the gas temperature from 3000~K to 6000~K. The CO outflow is found to be proportional to the plasma effective surface and exhibits no significant dependence on the actual \COO flow injected into the reactor but only on the input power at certain pressure. Thermal dissociation calculations show that, even at the lowest pressures of this study, the observed conversion and energy efficiency are compatible with a thermal dissociation mechanism.

\end{abstract}



\section{Introduction}

\label{sec:1.0} 
Carbon dioxide, a highly potent greenhouse gas, is produced in very large quantities from industrial processes but also for power generation. In May 2013, its concentration in the atmosphere exceeded 400~ppm in the atmosphere for the first time in modern history \cite{showstack2013carbon} which should be compared to the 280~ppm that characterized conditions before the industrial revolution \cite{monastersky2013global}. For achieving a target of \SI{2} {\celsius} maximum increase of Earth global warming, a complete decarbonization of the energy sector will be required by 2060 while scenario with maximum temperature of \SI{1.5} {\celsius} will require negative emission \cite{ETP2017}. Nowadays, the carbon dioxide emitted during electricity generation is contributing to about 20 \% of anthropogenic emission totalling 30 000 Mt/y \cite{Aresta2013} both from fossil and also biogenic raw materials. \COO is not currently consumed in large quantities by the industry (in 2011, it averaged to a mere 150 Mt/y worldwide) and generally it is treated as a disposable waste. In this perspective several strategy to increase the usage of \COO are being investigated \cite{Quadrelli2011}. 
Additionally, renewable energy sources suffer from intermittency and a significant geographical mismatch between availability and demand. Energy storage and transport are necessary in order to stabilize the power grid and match the consumers demand. Several groups around the world have proposed \COO as a primary building block for synthesis of synthetic fuels such as methanol \cite{osti_5529228}. \COO re-use for energy storage is an attractive option for a zero-emission carbon cycle while solving the issues of intermittency of renewable energies and their transport to remote locations. \COO can be used as raw material and building block for the production of CO and thereafter synthetic fuels \cite{Ganesh2014}.The current \COO valorization research focus mainly on electro-catalytic processes  and thermochemical processes \cite{schlogl2013solar}. More recently non-equilibrium plasmas attracted interest as possible means of \COO conversion, especially for energy storage purposes \cite{Bogaerts2018ACSEL}. Plasma devices offer high flexibility in terms of response times and scalability. 

There are different types of plasma discharges used for  \COO
conversion: dielectric barrier discharge, Microwave discharges,
radio-frequency discharges, corona discharges, gliding arc discharges, and nanosecond pulsed high voltage discharges. A comprehensive overview of the different devices used for \COO decomposition has been given by Snoeckx et al. \cite{Snoeckx2017}. Typically, DBD discharges have a maximum energy efficiency of 20\% and a \COO conversion of about 40\%. With gliding arcs on the other hand energy efficiency up to 60 \% have been reported but with conversion typically limited to a maximum of 30 \%. Microwave discharges and RF discharges are reported to achieve the highest conversion (up to 90 \%) and energy efficiencies (up to 80\%), albeit not simultaneously \cite{Snoeckx2017}. Microwave discharges represent the most promising technology for plasma assisted CO$_2$ conversion.

The dissociation of the CO$_2$ molecule requires a minimum energy of 5.5 eV, needed to break a \COO bond, but in presence of atomic oxygen additional dissociation of CO$_2$ can be obtained from the reaction \ref{eq:co2diss2}.

\begin{equation}
CO_2 + M \rightarrow CO + O + M  \:\:\:\:\:\:  \Delta H = 5.5 \: eV
\label{eq:co2diss1}
\end{equation}

\begin{equation}
CO^*_2 + O \rightarrow CO + O_2   \:\:\:\:\:\:  \Delta H = 0.3 \: eV
\label{eq:co2diss2}
\end{equation} 

The potential energy (2.6 eV) of the O atom produced in the first step (process \ref{eq:co2diss1}) is then not lost via its mutual recombination with another O atom to form an O$_2$ molecule via the reaction $O + O + M \rightarrow O_2 + M$ where M is a third particle or a wall, but rather invest to produce an additional CO molecule. In the latter scenario, the minimum energy requirement for one CO molecule production decreases to 2.93 eV/molecule. 

\begin{equation}
CO_2 \rightarrow CO + \frac{1}{2}O_2   \:\:\:\:\:\:  \Delta H = 2.93 \: eV
\label{eq:co2disscomplete}
\end{equation}

Thermodynamically, this reaction has a Specific Energy Requirement of 2.93~eV per dissociated CO$_2$ molecule at 400 K and at atmospheric pressure. The difference between process \ref{eq:co2disscomplete} and \ref{eq:co2diss1} is due to the potential energy difference of the O atom in its free or bonded state (i.e. O$_2$ molecule), which is equal to half the energy needed for splitting the O$_2$ bond. To obtain a specific energy requirement of 2.93 eV/molecule, it is then critical that the O atom also reacts with another CO$_2$ molecule to form a second CO molecule so that the potential energy of the O atom is not lost. 

In a plasma, the dissociation of CO$_2$ can be induced either by electron impact processes or thermal dissociation.  Thermal dissociation is the consequence of the shift of the chemical equilibrium due to a high gas temperature. Non-equilibrium plasmas (i.e. plasma with significantly lower gas temperatures compared to electron temperatures) can help to obtain energy efficiency higher then the once obtained by thermal dissociation of \COO, albeit with lower conversion rate. High energy electrons ($>$ 10 eV) can induce dissociation of the CO$_2$ molecule by exiting the molecule into a dissociative state \cite{Snoeckx2017}. The latter process requires overall more energy than the thermal dissociation, thus is unfavorable. Another electron driven process is the excitation of the asymmetric stretching of the CO$_2$ via several electron collision leading to the dissociation of the CO$_2$ molecule.  This mechanism is typically proposed in literature to explain an energy efficiency above 50 \% (thus higher then the thermal dissociation limit)  \cite{Fridman2008}. However such process requires a low gas temperature since the higher the gas temperature the faster are the losses of vibrational excitation into gas heating \cite{azizov1983Vakar}.

Due to their intrinsic low electric fields and high average power densities, microwave plasmas allow generating molecular plasmas with relatively low electron temperatures but high vibrational temperatures that can favor vibrational excitation in place of direct dissociation mechanisms by electron impact. Values up to 80 \% energy efficiency were reported by researchers from the Kurchatov Institute \cite{Rusanov1984, LEGASOV1978, ASISOV1983} and have not been reproduced in recent experiments. The maximum energy efficiency was observed at pressures around 200 mbar \cite{Butylkin1981} and degraded at higher and lower pressures.
Nowadays, two types of microwave sources are mainly used to investigate \COO dissociation: in 915~MHz  sources values of energy efficiencies up to 50 \% and conversion up to 80 \% have been obtained, although not simultaneously \cite{Bongers2017}. Typically energy efficiency and conversion anti-correlates,an high value of conversion implies low energy efficiency and \textit{viceversa}. In the 2.45~GHz sources the energy efficiency is between few percent and 30~\% are obtained \cite{Belov2018}, \cite{Bekerom2018}, \cite{Bongers2017}, \cite{Rooij2015}. Belov et. al. \cite{Belov2018} investigated a 2.45 GHz microwave plasma torch in a pressure range between 200 mbar and 900 mbar, with variable flow injection geometry (i.e. direct, tangential and inverse), concluding that the flow dynamics strongly influences the overall \COO conversion. Recent work by van den Bekerom et al. \cite{Bekerom2018} have shown that most of the literature results do not outperform the best possible thermal efficiencies and that in microwave plasma thermal dissociation and the plasma volume (heated fraction) leads to much higher effective SEI locally in the plasma.

To the best of our knowledge, no setup has been investigated systematically in both large pressure and gas flow range. This work focuses on the characterization of a 2.45 GHz microwave plasma torch in a wide flow power and pressure range as means of CO$_2$ conversion into CO, focusing on correlating the observed energy efficiency and conversion with the gas temperature and the plasma volume.

\section{Experimental setup}
\label{sec:2.0} 
\begin{figure}[H]
\centering
\subfloat[][\emph{Mass spectrometry configuration}]{\includegraphics[width=0.48\textwidth]{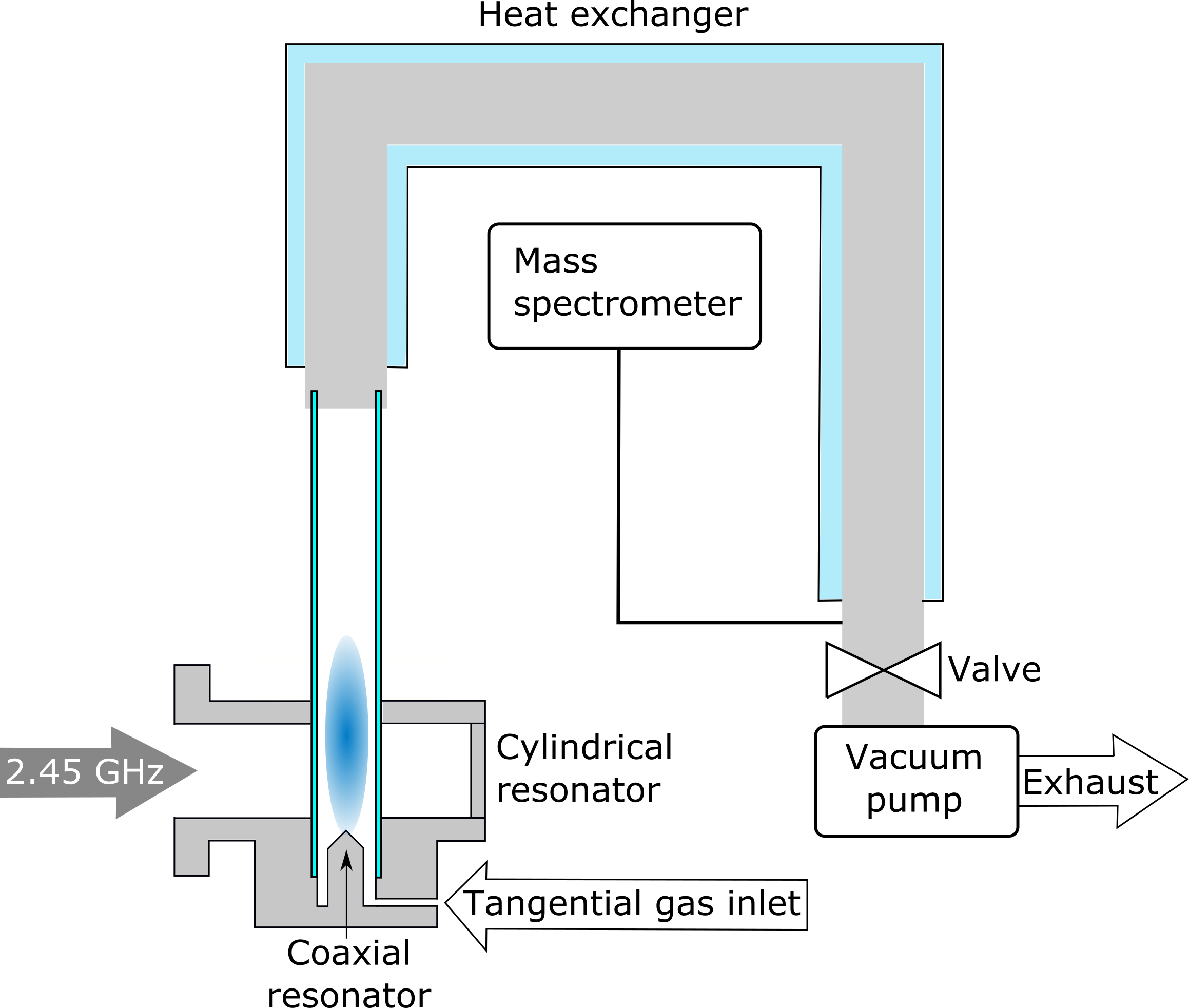}}
\subfloat[][\emph{iCCD imgaing configuration}]{\includegraphics[width=0.57\textwidth]{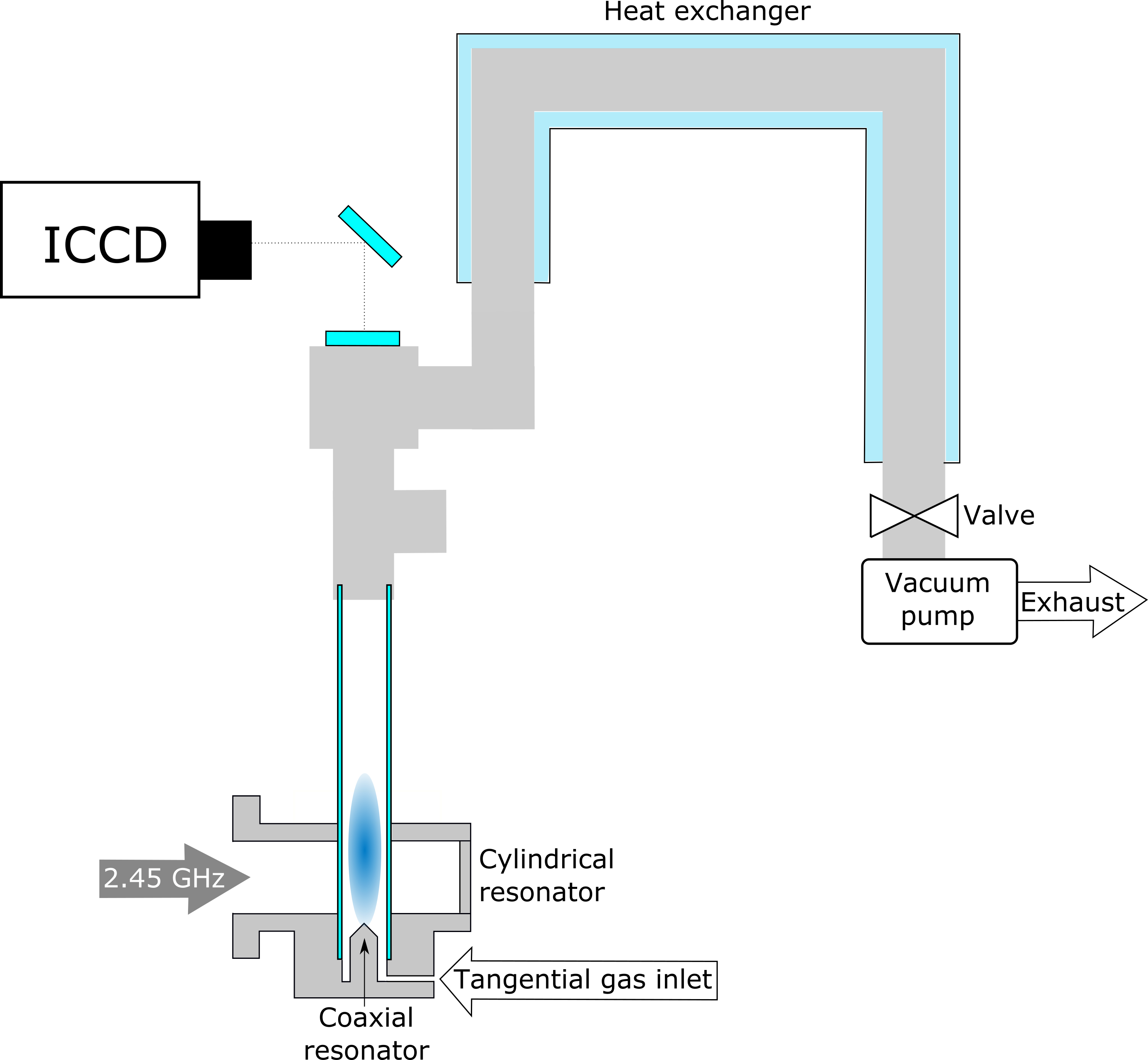}}

\caption{Schematic of the plasma torch and its exhaust system. Figure  \textit{(b)} shows the modified setup to measure the axial integrated optical emission. The most important components are indicated in the scheme.}
\label{fig:expsetup}
\end{figure}

The plasma torch used in this work has been modified from the original design of University of Stuttgart \cite{Leins2013} so that it can be operated at atmospheric pressure but also at lower pressures.  The experimental setup used in this work is shown in figure \ref{fig:expsetup} \textit{(a)}, it consists of a cylindrical TE$_{10}$ cavity and a coaxial resonator. The coaxial resonator placed at the bottom of the quartz tube consists of two elements: a $\lambda$/4  resonator and a tip in its center. The tip  consists of two geometrical parts: a cylinder of height 12 mm with a diameter of 15 mm and a cone of 8 mm height with a base diameter of 15 mm. The tip can be adjusted in height to enhance the electric field at its top for a given microwave frequency  \cite{PhDLeins}. The tip position has been adjusted to have an enhanced electric field at 2.45~GHz by using a network analyzer. The enhanced electric field allows ignition in the pressure range 10-1000~mbar. A quartz tube of 30 mm outer diameter and 26 mm inner diameter is mounted in the center of the cylindrical resonator as can be seen in figure  \ref{fig:expsetup} \textit{(a)}. The quartz tube length can be varied from a minimum of \ca 8~cm up to 40~cm, in this work a 40~cm has been used. The cylindrical cavity has 3 vertical slits of width 5.5 mm and height 43 mm (that corresponds to the cylindrical resonator height) which allow optical access to the plasma in the resonator. At the bottom of the coaxial resonator the gas is injected in the quartz tube by four tangential gas inlets of 4.3 mm diameter. The 2.45 GHz microwave are generated by a magnetron MH3000S-213BB powered by a 3 kW power supply ML3000D-111TC, both are Muegge GmbH components. The power supply can be operated with a microwave power output that ranges between 300 W and 3000 W. The plasma is ignited and confined in the quartz tube.  The end of the quartz tube is connected to a 2~m long  water cooled pipe. The plasma effluent is pumped away by a vacuum pump with variable pumping speed. The system is operated at pressure between 30 mbar and 1000 mbar with CO$_2$ flow rate between 2 and 100 L/min. However a \COO flow below 5 L/min typically generate unstable plasma at pressure above 100 mbar. At pressure lower than 100 mbar to use a low \COO flow typically result in quartz tube heating when a microwave power above 1 kW is coupled into the plasma. The present setup differs from the one used in the Differ institute for the presence of the ignition pin \cite{Bongers2017}. Moreover excepted for the inlet gas velocity and a slightly different microwave frequency (2.4 GHz), that also implies differences in the size of the microwave components, the present setup is similar to the one described by Butylkin \cite{Butylkin1981} for which energy efficiencies up to 80 \% were reported. The plasma obtained can be operated in a flow range between 5 and 100 L/min.

Figure \ref{fig:expsetup} \textit{(b)} shows the modifications of the experimental setup for measuring the plasma emission integrated along the axial direction (i.e. gas stream direction) and determining its radial cross section: a CF40 cube is introduced on top of the plasma torch, the cube side that face the quartz tube is provided with a quartz Thorlabs window WG42012. A Thorlabs mirror PF20-03-01 reflects the emitted light that is than measured with an ANDOR iStar ICCD camera with 2048x512 pixels of 13 $\mu$m size and a squared intensifier of side 18 mm, equipped with a Nikon lens of 35mm focal length.

Optical emission spectroscopy is performed on the light collected via a double iris system that reduces the collection area to about 1 mm$^2$. The collected light is analyzed using a SPEX-1000 spectrometer (with a 1800 l/mm grating) provided with an Andor AK420-OE CCD camera. The optics have been absolutely calibrated in the wavelength range in the range between 250 nm to 400 nm with a Deuterium lamp and between 400 nm to 850~nm with an Ulbricht sphere (Labsphere USS-800C-100R in combination with a LPS-100 power supply).  

Figures  \ref{fig:expsetup} \textit{(a)} shows the position where the plasma effluent is analyzed with the mass spectrometer. The gas is sampled after the heat exchanger about 2 m after the end of the plasma using a 0.9 mm inner diameter tube. To reduce the pressure from atmospheric to the mass spectrometer working pressure (10$^{-6}$ mbar) two orifices are used. A 100 $\mu$m orifice connects the 0.9 mm sampling tube with an intermediate T-shaped chamber that is kept at the constant pressure of 1 mbar. The T-shaped chamber is connected to the ionization chamber of the mass spectrometer via a variable orifice that allow to control the pressure in the ionization chamber. The mass spectrometer has been calibrated to measure the concentration of CO, CO$_2$, O$_2$ using known gas mixtures. The analysis of the system behavior shows that the calibration is valid in the pressure range investigated and no significant gas de-mixing takes place \cite{MSpaper}. The determination of the gas composition has been carried out by calculating a synthetic mass spectrum and using a least square minimization routine (python library scypy.optimize method L-BFGS-B) on the experimental data in order to determine the particle concentration in the gas \cite{Drenik2017}. A synthetic mass spectrum is calculated as 
\begin{equation}
I(m)=  \sum_{i}{ \beta_i n_i c_i^{m}}
\label{eq:MSPeak}
\end{equation}
where I(m) is the peak intensity at mass m, $\beta_i$ the MS calibration parameter for the specie i,  n$_i$ is the species relative density of species i and c$_i^{m}$ is the contribution (cracking pattern) to mass m of the species i. The conversion rate of our plasma can be determined by analyzing the gas composition as
\begin{equation}
\chi =  1 - \frac{[CO_2]_{out}}{[CO_2]_{in}} = \frac{1- \frac{[CO_2]_{out}}{[CO_2]_{out} + [CO]_{out} +[O_2]_{out}}} {1 - \frac{[CO_2]_{out}}{2\cdot \left([CO_2]_{out} + [CO]_{out} +[O_2]_{out}\right)}}
\label{eq:conv}
\end{equation}
   
where $\chi$ is the conversion efficiency, [CO] and [CO$_2$] stand for the carbon monoxide and carbon dioxide concentration. The more common form of the conversion  $ \frac{[CO]_{out}}{[CO_2]_{out} + [CO]_{out} }  $ can be obtained under the assumption of 2~[CO]$_{out}$~=~[O$_2$]$_{out}$, fulfilling the stoichiometry of equation \ref{eq:co2disscomplete}. The latter is not assumed \textit{a priori} equal to 2, but rather used as control parameter for the correctness of measurements performed as will be discussed by Hecimovic et al. \cite{MSpaper}. From the conversion efficiency the energy efficiency can be calculated:  

\begin{equation}
\eta = \chi \frac{\Delta H}{SEI}
\label{eq:energy}
\end{equation}

$\eta$ is the energy efficiency, $\Delta$H the enthalpy of the CO$_2$ dissociation (2.93 eV), and SEI is the global specific energy input calculated as:

\begin{equation}
SEI=  0.0138 \: \frac{power[W]}{flow[L/min]}  \bigl[\frac{eV}{molecule}\bigr]
\label{eq:SEI}
\end{equation}
where the power is expressed in Watt and the flow in standard liter per minute, the overall constant has the proper units to obtain the SEI in eV/molecule \cite{Britun2018}. The total flow is used to calculate SEI and not the one that effectively interact with the plasma. 

\section{Experimental results}
\label{sec:results}


\subsection{$\mathrm{CO_2}$ conversion and energy efficiency}
\label{sec:3.1} 

\begin{figure}[H]
\centering
\subfloat[][\emph{Quasi-atmospheric pressure}]{\includegraphics[width=0.48\textwidth]{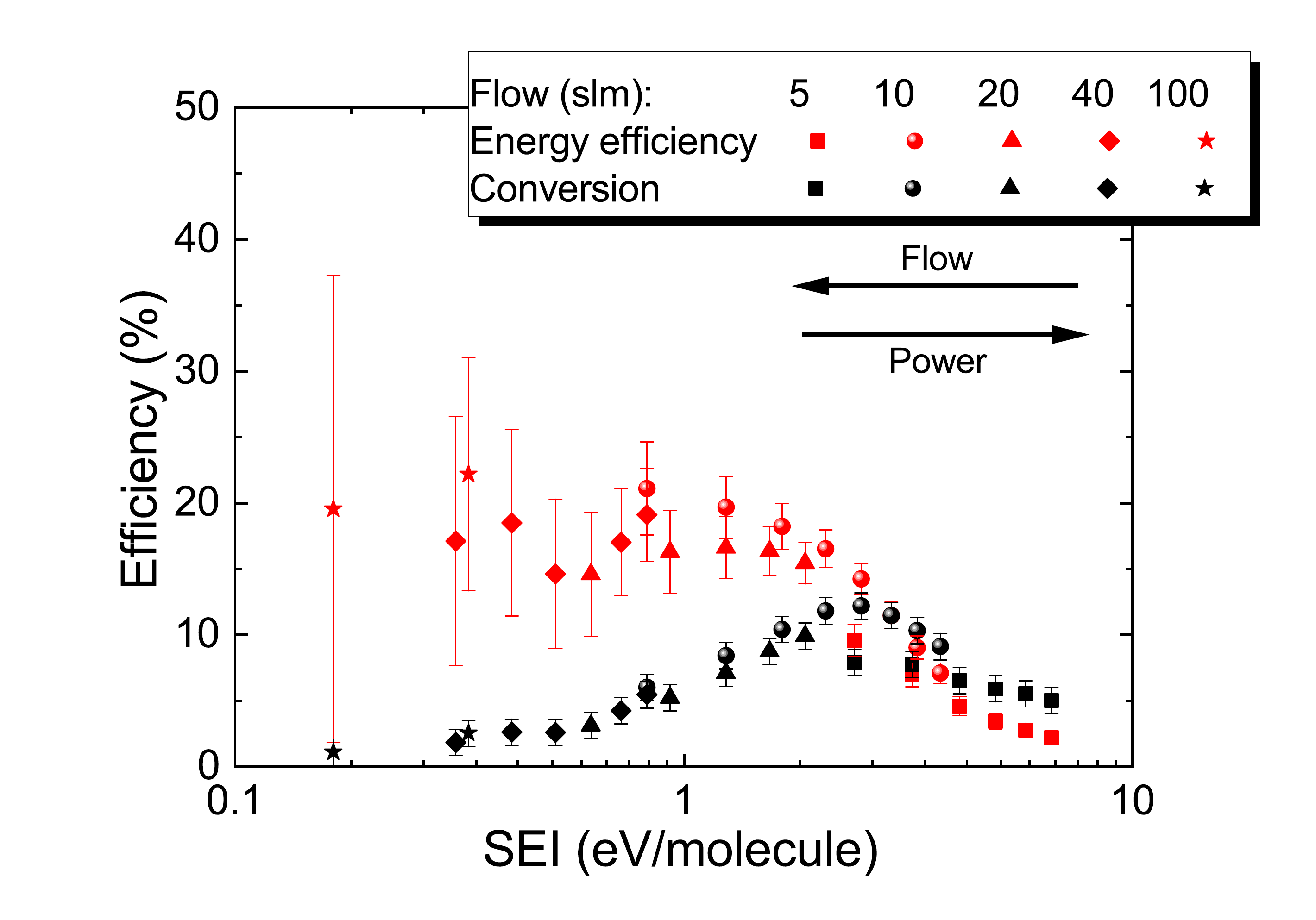}}
\subfloat[][\emph{500 mbar}]{\includegraphics[width=0.48\textwidth]{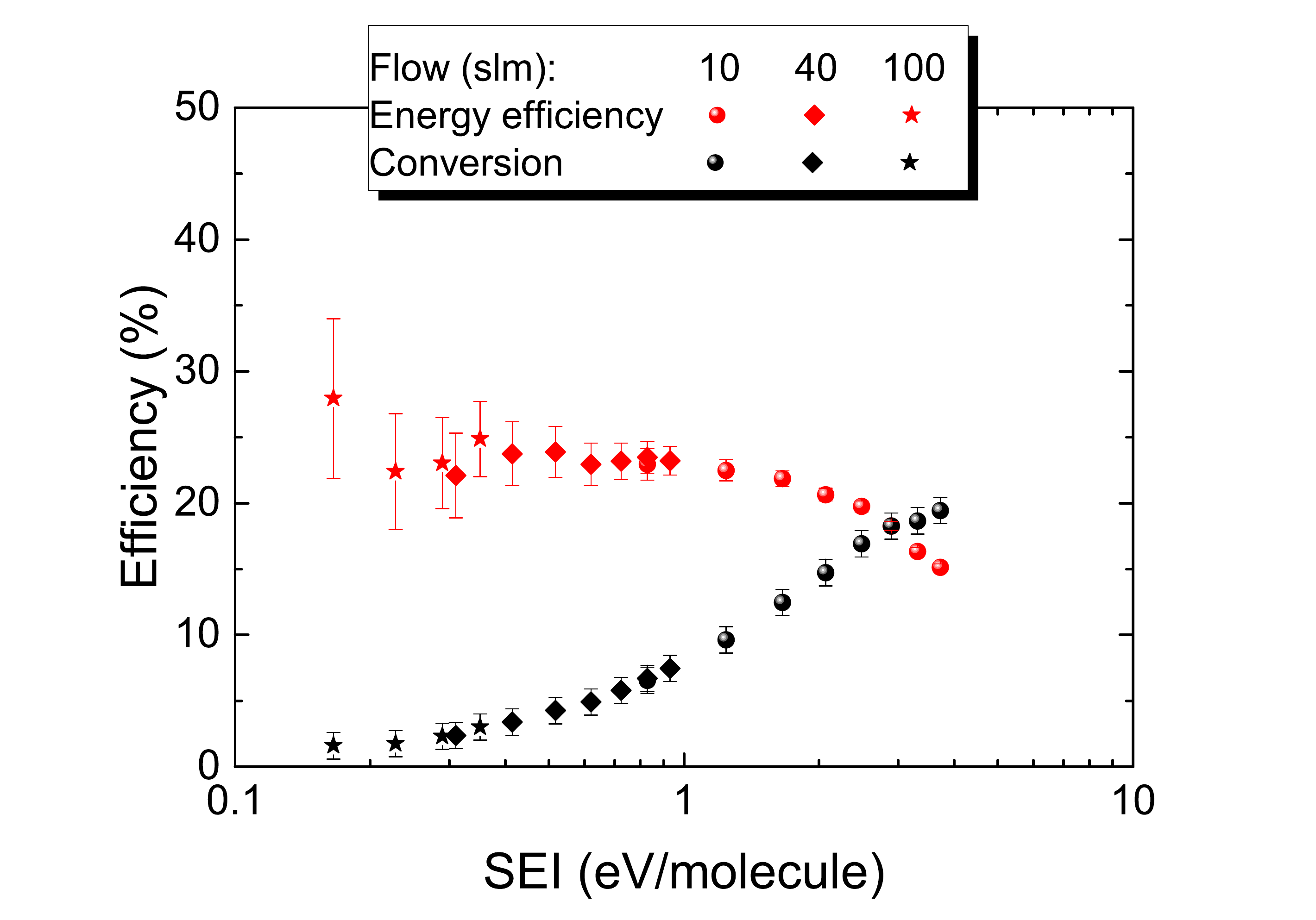}}

\subfloat[][\emph{200 mbar}]{\includegraphics[width=0.48\textwidth]{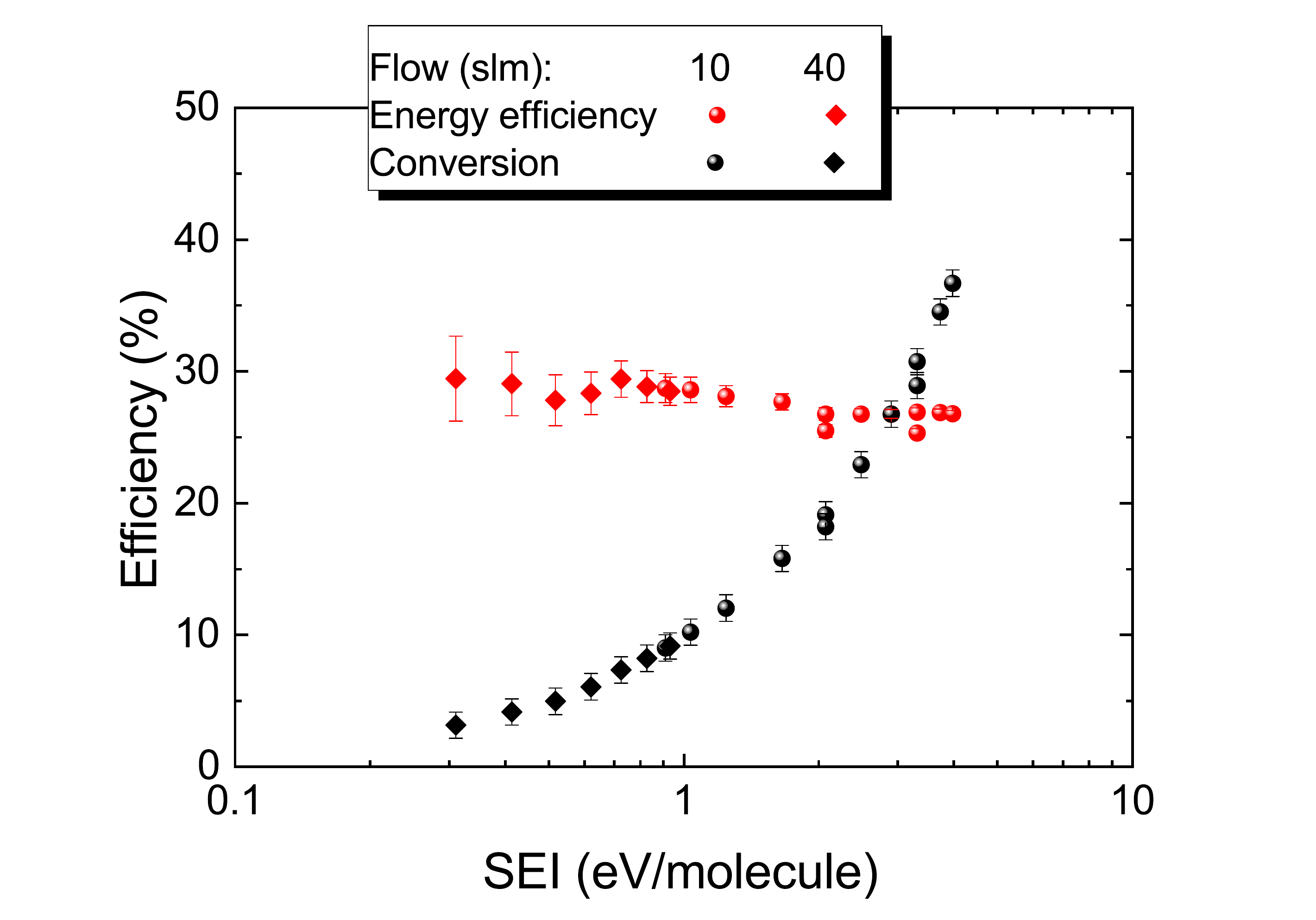}}
\subfloat[][\emph{60 mbar}]{\includegraphics[width=0.48\textwidth]{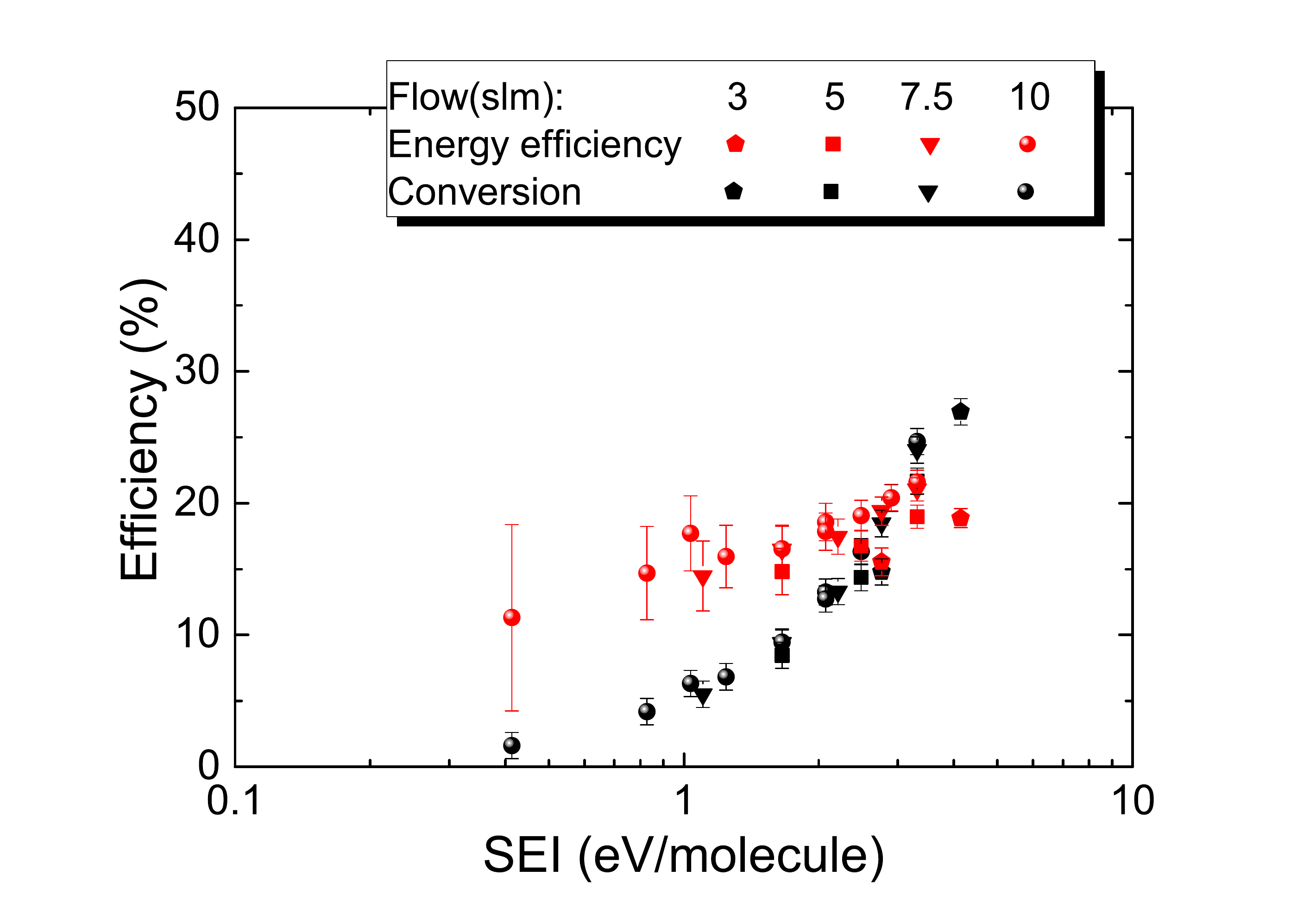}}

\caption{Energy efficiency (red dots) and conversion (black dots) as function of the SEI. Flow and power have been varied between 3 and 100 L/min (depending on the pump capacity) and between 900 and 2700 W, respectively. The effect of \COO flow and power on the SEI is indicated by the arrows in figure \fa.  
}
\label{fig:MS}
\end{figure}

The capability of the plasma torch for converting CO$_2$ into CO and its energy efficiency is measured in a wide pressure range using mass spectrometry. The source performance from 60 mbar to quasi-atmospheric pressure (i.e. 880-930 mbar; the precision was limited by mechanical precision of the regulation valve) is shown in figure \ref{fig:MS}. The energy efficiency and conversion are shown as function of the global specific energy input (cf. equation \ref{eq:SEI}). The measurement of the CO$_2$ conversion is known within $\pm$~1~\%, because of systematic errors in the calibration procedure and background subtraction. The error on the energy efficiency is obtained by propagating the uncertainty on the conversion. 


Figures \ref{fig:MS} \fa, \fb, \fc, \fd $\:$ show that the amount of CO$_2$ converted typically depends on the SEI: the higher the energy per molecule, the higher the conversion. At 60~mbar (figure~\fd) the conversion increases from values of \ca 1~\% at SEI \ca 0.4~eV/molecule up to \ca 30~\% at SEI \ca~4~eV/molecule. At 200 mbar (figure~\fc) the increase obtained by increasing the SEI is stronger: a conversion of \ca 3~\% is measured at \ca 0.3~eV/molecule and a conversion of \ca 35~\% at \ca 4~eV/molecule. 
At 500 mbar (figure~\fb) the conversion increases from \ca 2~\% at \ca 0.2~eV/molecule up to 20~\% at \ca 4~eV/molecule.  A remarkable deviation from the established trends can be observed at SEI above 2 eV/molecule at 500~mbar 
the conversion saturates increase of power (i.e. SEI at fixed flow) does not produce a further increase of the fraction of CO$_2$ converted. At quasi-atmospheric pressure (figure~\fa) an increase of energy efficiency produces a decrease of conversion above 2~eV/molecule. In the latter case the conversion increases from \ca 1\% at \ca 0.2~eV/molecule to \ca 13~\% at \ca 2~eV/molecule and then decreases to 5~\% at \ca 8~eV/molecule. The maximum  conversion that can be achieved in the present setup is depending on the pressure, the lower the pressure the higher conversion can be achieved (by increasing the power).
The energy efficiency is weakly changing with the SEI and is observed (almost) constant within error-bars in most of the conditions studied as can be observed in figures \ref{fig:MS} \fc, \fd and in figures \ref{fig:MS} \fb, \fa at SEI below 2 eV/molecule. The energy efficiency changes with pressure, the maximum of \ca 30 \% is measured at 200 mbar. No flow effect onto the energy efficiency can be detected, as opposed to what has been previously observed at pressures between 200 mbar and atmospheric pressure by Belov et al. \cite{Belov2018}. They reported an increase of the energy efficiency while using larger gas flows. However it should be noted that the experimental setup used in this work is different from the once used by Belov et al. \cite{Belov2018}, particularly in terms of plasma cross section (140~mm) and although a vortex injection configurations was use, its geometry is different from the one used in this work, these differences can be the origin of the discrepancy. 

The conversion and energy efficiency are typically depending only on the SEI, but at quasi-atmospheric pressure with a \COO flow of 5~L/min the observed conversion deviates from the trend observed at 10~L/min. The lower flow is (probably) the main responsible of the observed deviation, but the physical mechanism behind it is still unclear. Such trend could also be present at lower pressures, but shifted to higher SEI values. The latter hypothesis is supported by figure \ref{fig:MS}~\fb where a decrease in conversion in observed at SEI of \ca 2~eV/molecule. Further investigation in this direction are needed. Nevertheless carrying out measurements at SEI above 4~eV/molecule at pressure above 200~mbar is challenging, because of limitation in the MW power (max. 3000~W) and since the plasma is unstable at flows below 5 slm and pressures above 200~mbar.

\begin{figure}[H]
\centering
\subfloat[][\emph{}]{\includegraphics[width=0.48\textwidth]{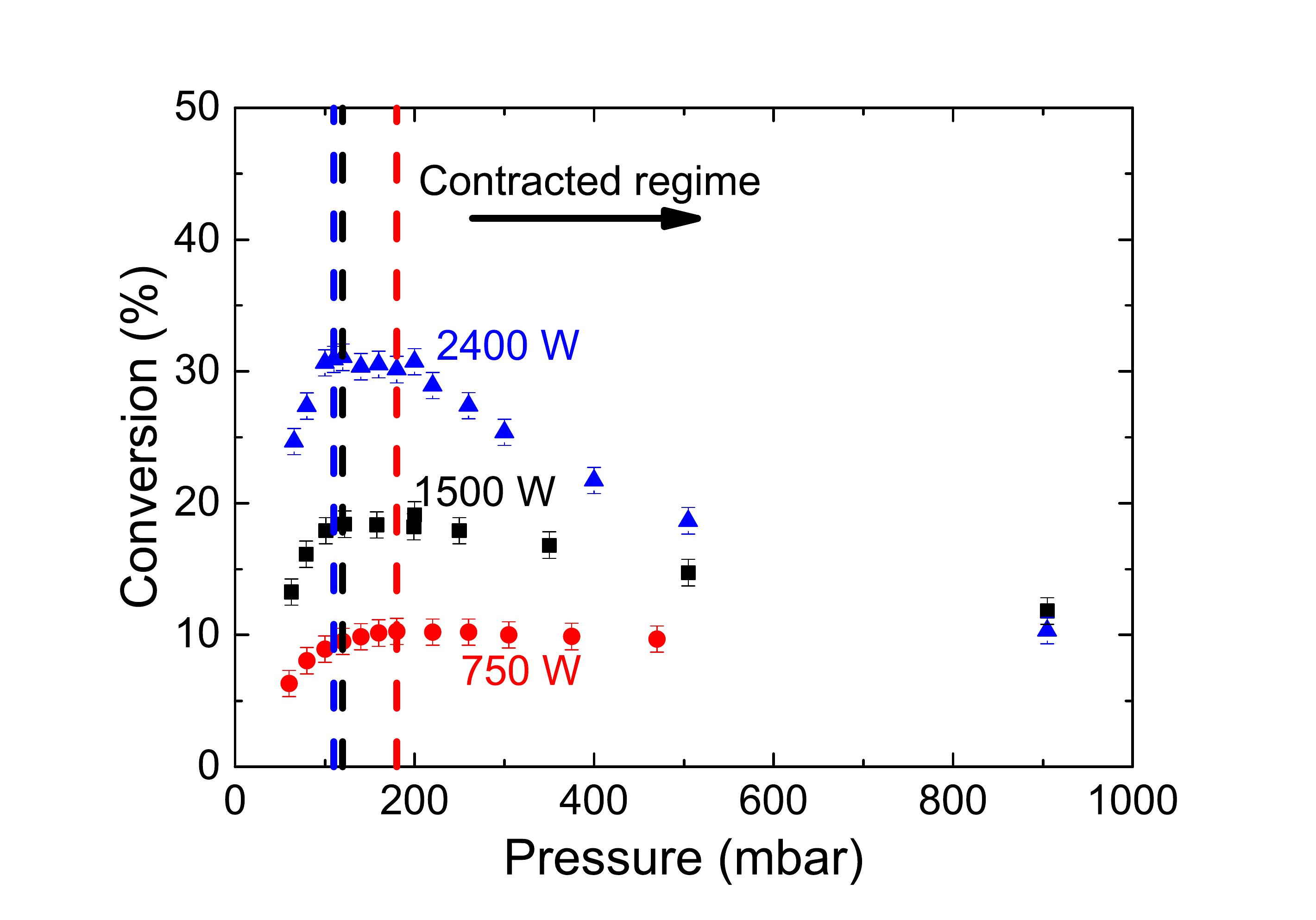}}
\subfloat[][\emph{}]{\includegraphics[width=0.48\textwidth]{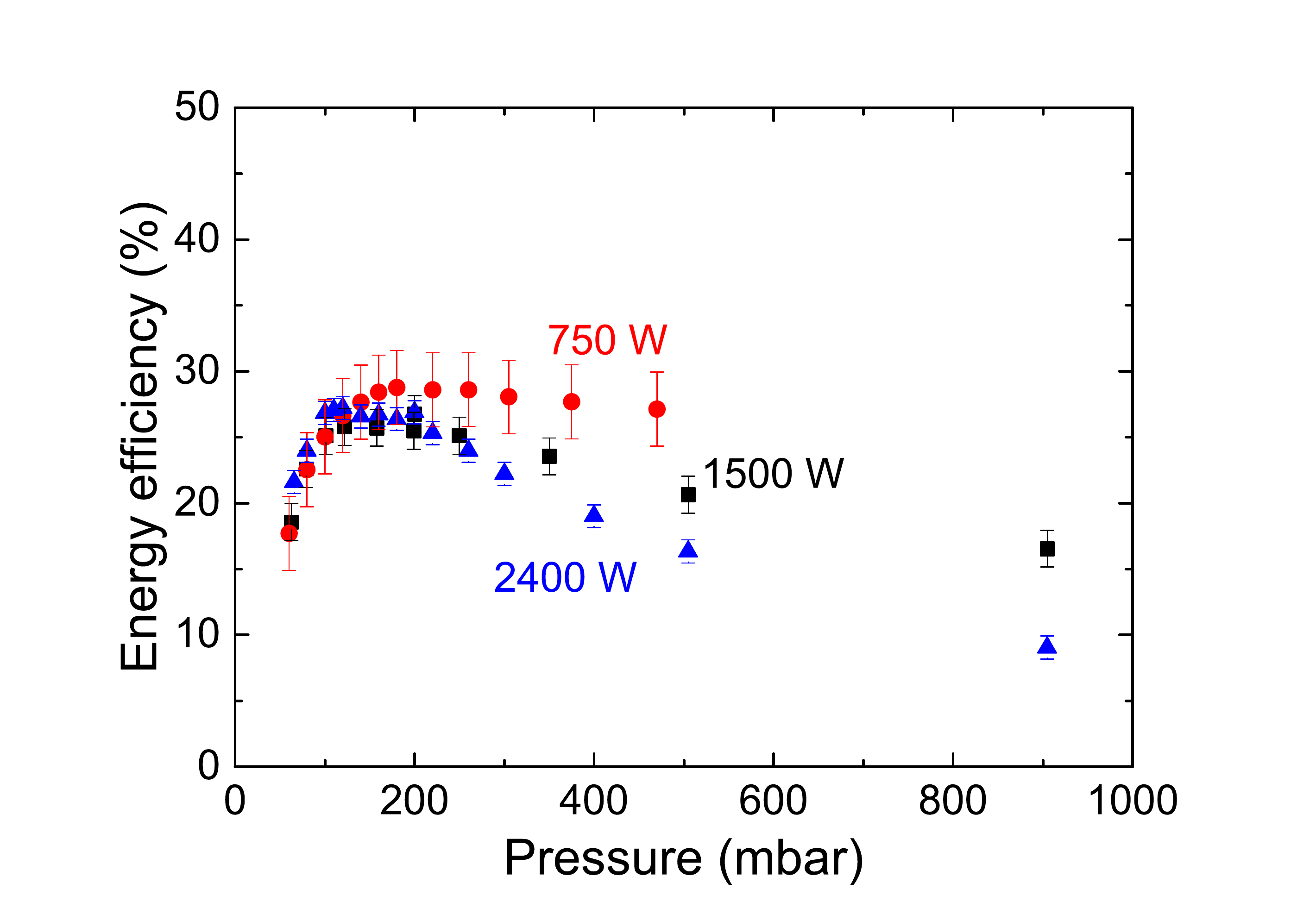}}

\caption{Figures \textit{(a)} and \textit{(b)} show the conversion and energy efficiency as function of pressure at constant flow 10 L/min at three different power 750 W, 1500 W and 2400 W represented in red, black and blue respectively. In Figures \textit{(a)} the dashed lines mark the pressure at which the plasma contracts at the studied powers.}
\label{fig:pscan}

\end{figure}

The conversion of \COO  into CO is investigated as function of the pressure at fixed flow and power and the results are shown in figure \ref{fig:pscan} \textit{(a)} and \textit{(b)} respectively. A contraction of the plasma takes place (see section \ref{sec:3.3} for more details) at about 120~mbar with the exact pressure value depending on the power coupled to the plasma: at 750 W the plasma contracts at 180~mbar, at 1500~W at 120~mbar and at 2400~W at 110 mbar. In figure \ref{fig:pscan} the contraction pressure is indicated by vertical dashed lines. The amount of \COO dissociated increases until the plasma contracts. The energy efficiency before the plasma contraction increases rapidly with pressure, it reaches an optimum at pressures near the contraction and then stays constant at low power and decreases at higher power. This observation is in agreement with Fridman overview and discussion of results from the Kurchatov institute \cite{Fridman2008}. After the contraction takes place the conversion of \COO reduces with increasing the pressure, with the trend of the higher the power the stronger is the reduction. The energy efficiency follows similar trend as conversion. Belov et al. \cite{Belov2018} also reported improved energy efficiencies from atmospheric pressure down to 200~mbar for a different microwave plasma source but they were not able to measure at lower pressure.   


\subsection{Gas temperatures of the plasma}
\label{sec:3.2} 
\begin{figure}[H]
\centering
\includegraphics[width=0.95\textwidth]{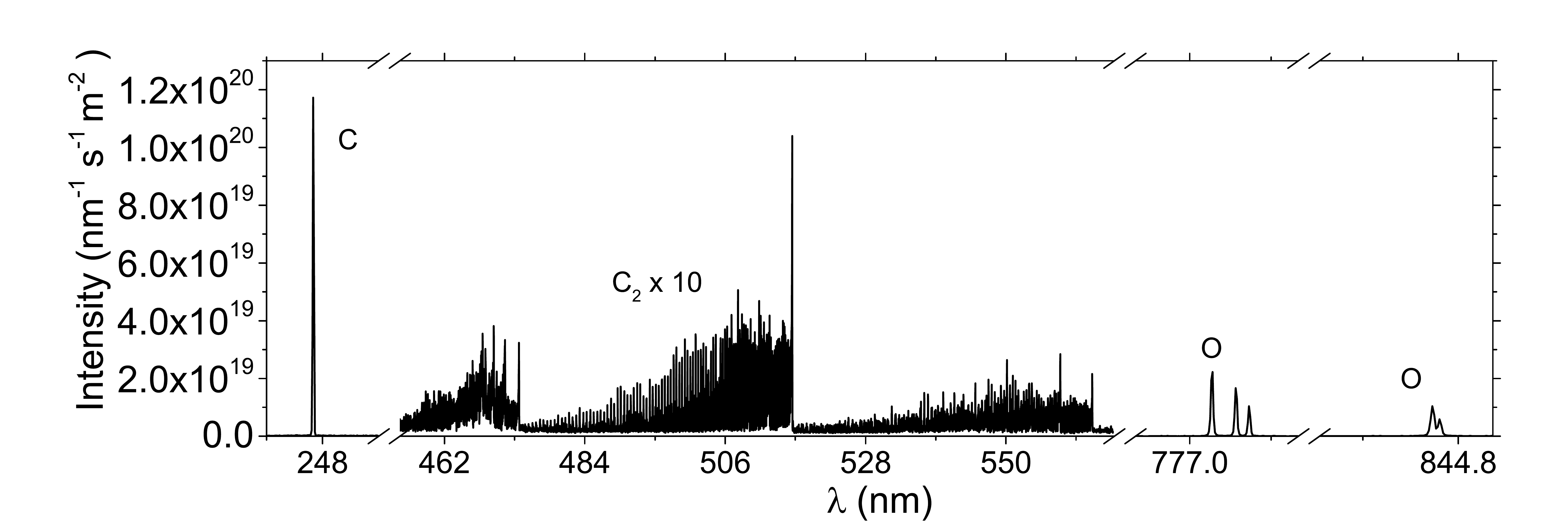}
\caption{Typical emission spectra absolutely calibrated recorded at 920 mbar, 10 L/min and 900 W. The emission was recorded in the center of the resonator. The spectra region between 460 nm and 567 nm (C$_2$ emission) has been multiplied by 10.}
\label{fig:C2spectra}
\end{figure}

\begin{figure}[H]
\centering
\includegraphics[width=0.95\textwidth]{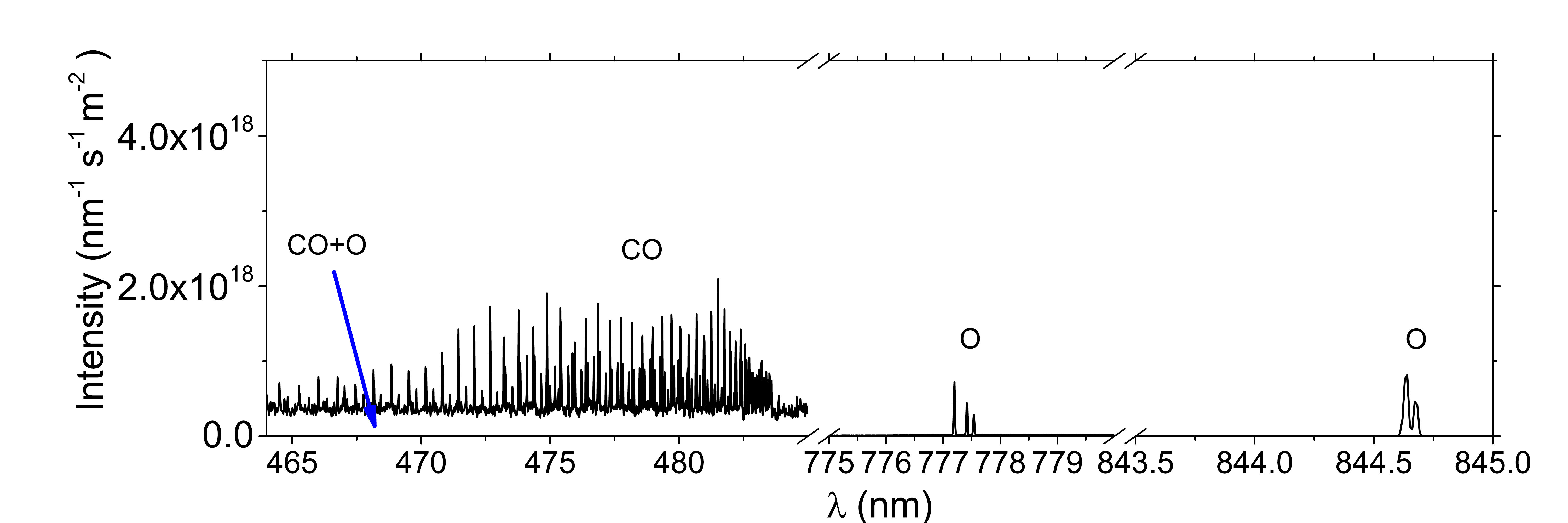}
\caption{Typical emission spectra absolutely calibrated recorded at 60 mbar, 10 L/min and 900 W. The emission was recorded in the center of the resonator.}
\label{fig:COspectra}
\end{figure}

To get more insight into the mechanisms of CO$_2$ dissociation inside the plasma, the gas temperature is studied by means of optical emission spectroscopy. Figure \ref{fig:C2spectra} shows a typical emission spectrum, recorded in the resonator, of the plasma operated in the contracted regime. It is dominated by the C$_2$ Swan (\Swan) bands (visible in the range between 460 nm and 570 nm) similarly to what has been previously observed \cite{Mitsingas2016}, \cite{Spencer2012}, \cite{Bongers2017}, \cite{Babou2008}. A broadband continuum emission extending in the range 300-700~nm is also always present. Such emission is typically (much) less intense than the C$_2$ Swan band or the atomic lines, but its contribution can be distinguished in the baseline of high resolution spectra (see figure \ref{fig:COspectra}) or in low resolution spectra since the lower dispersion allows better signal to noise ratio. Its origin is attributed to recombination processes of oxygen atoms via O + O $\rightarrow$ O$_2$ + h$\nu$ and CO + O + M $\rightarrow$ CO$_2$ + M + h$\nu$ reactions \cite{FloranISPC24}. In addition to that, one can identify some typical carbon (248 nm) and oxygen (at 777 nm and 844 nm) neutral lines. 

Figure \ref{fig:COspectra} shows the typical emission in the expanded regime: the spectrum is dominated by the continuum emission, on top of which can be identified the CO Angstrom (\Angstrom) bands. The atomic oxygen lines can be observed at 777~nm and 844~nm, but no atomic carbon lines are detected. In the expanded regime that is observed only at low pressures, no C$_2$ Swan bands are observed.

 The C$_2$ Swand bands emission spectra are fitted with synthetic spectra calculated using massiveOES \cite{MassiveOES} in combination with a database of lines and transition probabilities from Brooke et al. \cite{BROOKE}. The fitted temperature assuming a Boltzmann distribution allows describing accurately the rotational and vibrational population distributions (i.e no deviation from Boltzmann distributions is found) which can therefore be associated to the gas temperature of the plasma (see Carbone et al \cite{C2paper} for more details). The analysis of the C$_2$ Swan band shows consistently a gas temperature of about 6000 K $\pm$ 500 K in the center of the quartz tube independently of plasma conditions in the contracted mode (see section \ref{sec:thermo} for a discussion on that point). In the investigation of optical emission of C$_2$ in \COO microwave plasma carried out by Carbone et al. on the same plasma setup was shown that the measured rotational and vibrational temperatures are constant within error-bars and equal to each other in the constricted regime with a value of 6000 K $\pm$ 500 K. A parametric study was performed while varying the power between 900 W and 3 kW and using 5-100 L/min input gas flow \cite {C2paper} (i.e. similar conditions as the one investigated in sections \ref{sec:3.1} and \ref{sec:3.3}). 

\begin{figure}[H]
\centering
\subfloat[][\emph{}]{\includegraphics[width=0.48\textwidth]{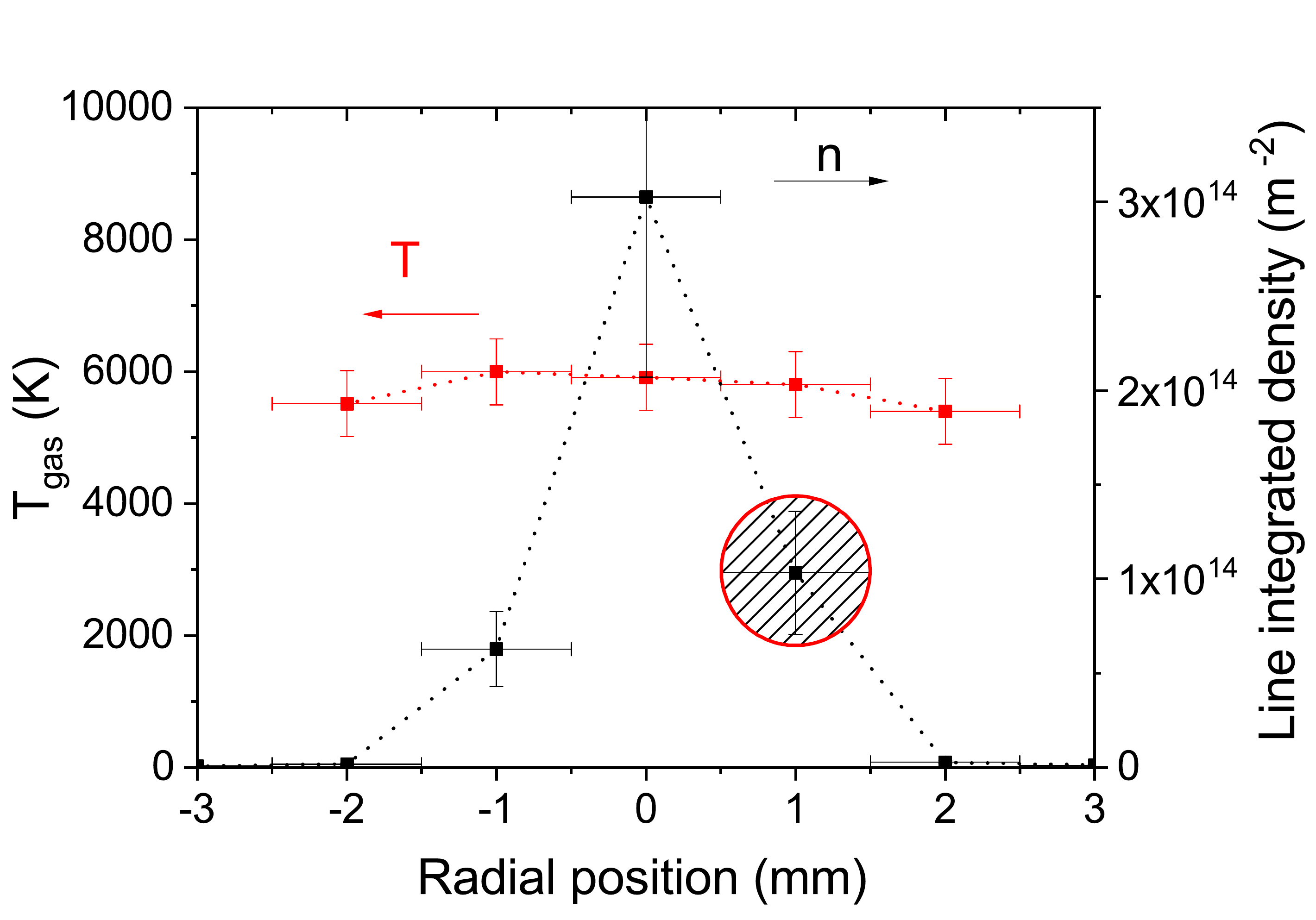}}
\subfloat[][\emph{}]{\includegraphics[width=0.48\textwidth]{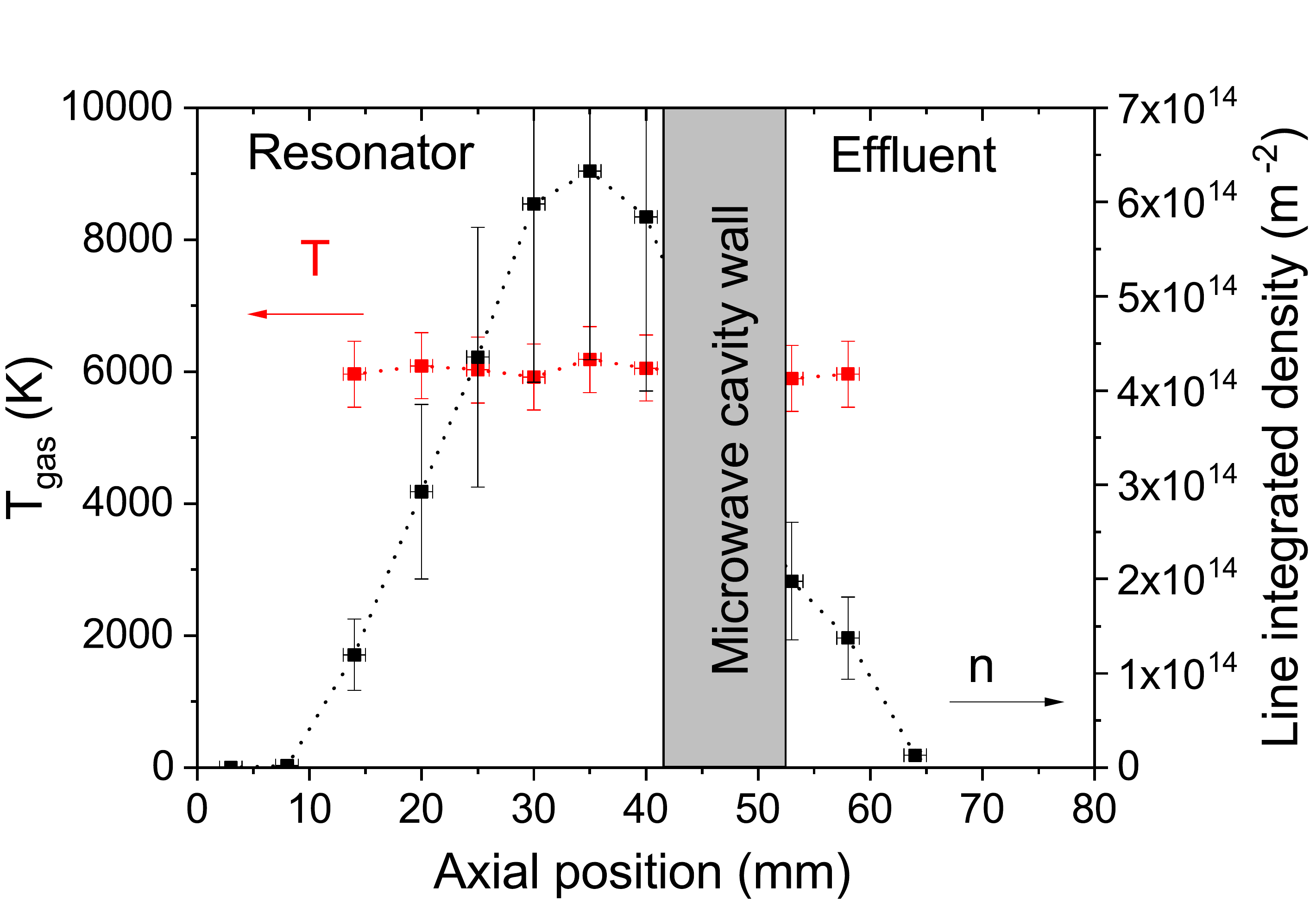}}

\caption{Figure \textit{(a)} shows the temperature evolution and the line integrated population density of the C$_2$ (d$^3\Pi_g$) state in the radial direction, red and black dots respectively. The measurements where performed at height of 58 mm  from the resonator bottom. Figure \textit{(b)} shows the temperature evolution and the population density of the C$_2$ (d$^3\Pi_g$) in the axial direction, red and black dots respectively. The measurements are performed in the center of the quartz tube, r=0. In both cases the CO$_2$  flow is 10 L/min, the pressure 920 $\pm$ 10 mbar.  The analysis of the C$_2$ emission has been performed on the $\Delta \nu$ = 0 transition group.}
\label{fig:axialradialscan}
\end{figure}

Figures \ref{fig:axialradialscan} \textit{(a)} and \textit{(b)} show the typical gas temperature and line integrated particle density evolution along the plasma radial and axial direction at 920~mbar. The LOS area is \ca 1~mm$^2$ (see section \ref{sec:2.0}). Since the double iris system is moved by a $\mu$m translator whose precision is much smaller than the LOS size, the precision on the measurement position is assumed to be \ca $\pm$0.5~mm. The measurements are performed with a CO$_2$  flow of 10 L/min, a pressure of 920~mbar and microwave power of 900 W. The radial analysis is performed at a fixed height of 58 mm, thus in the early effluent. The axial measurements are taken along the axis of the quartz tube inside the microwave resonator. In both cases the gas temperature measured from the C$_2$ rotational population distribution is constant at about 6000 $\pm$ 500 K. The axial analysis shows that the C$_2$ emission peaks in the resonator upper part and decay in the effluent. The upward shift of the plasma in the resonator is probably due to a combination of effects related to the gas flow and the distribution of the electromagnetic field inside the cylindrical cavity. The radial scan shows that the hot region  (where C$_2$ emits) occupies only a small portion of the quartz tube. This is in accordance with the measurements reported in section \ref{sec:3.3}. No temperature gradient are measured in the radial direction but this can be explained by the fact that only a small region is probed. Indeed, the C$_2$ molecules emit only in the core of the plasma where the gas temperature is the highest. Formation process of C$_2$ involves carbon atoms (as discussed by Carbone et al. \cite{C2paper}) that can form (thermally) only at temperature above \ca 5000~K (see section \ref{sec:thermo}). Groen et al. \cite{Groen2019} reported similar values and profile by O line Doppler broadening measurements in a pure CO$_2$ microwave discharge in constricted regime.

\begin{figure}[H]
\centering
\includegraphics[width=0.55\textwidth]{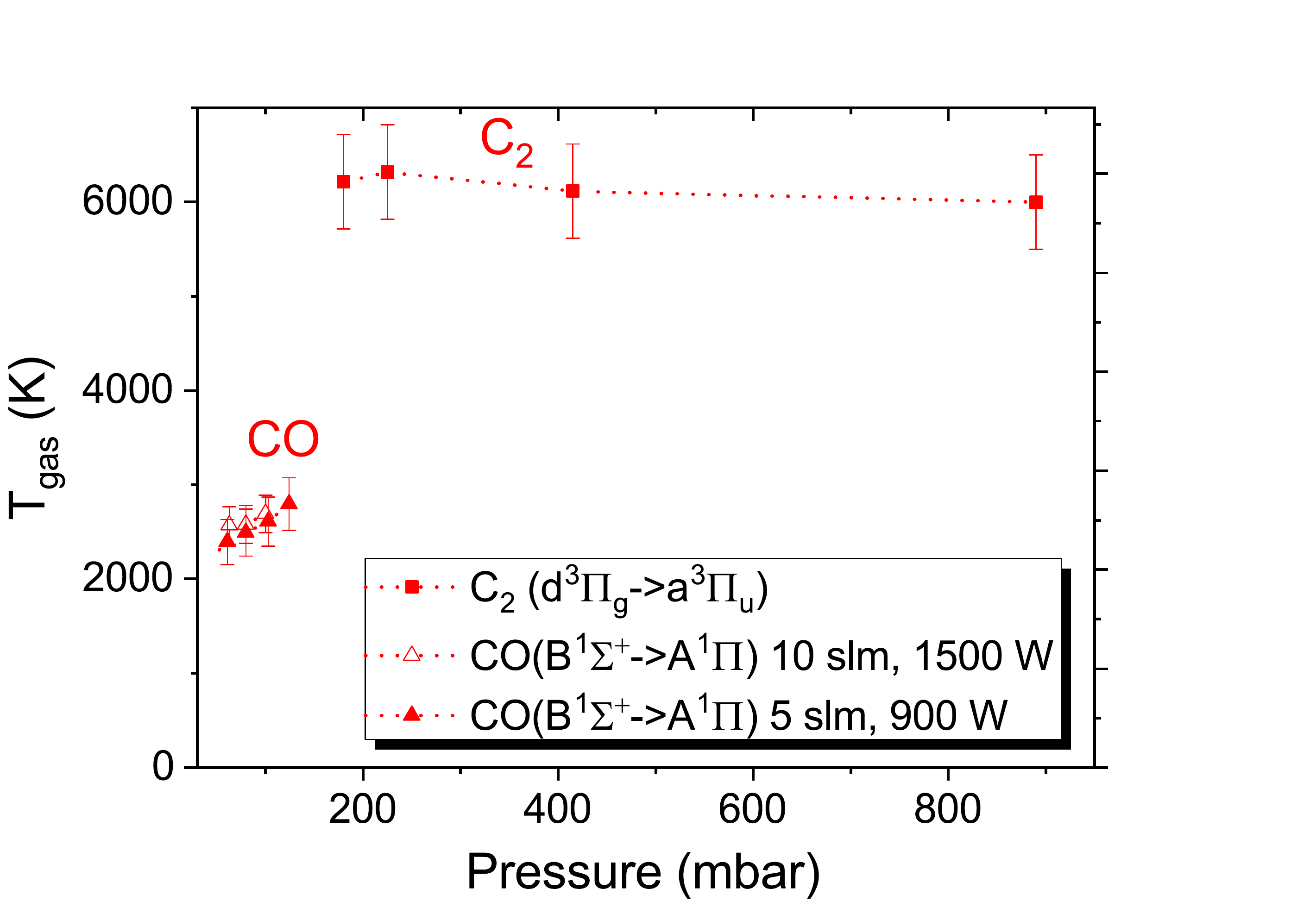}
\caption{Gas temperature (=T$_{rot}$)  as function of pressure is represented by the red triangles and red squares. The red triangles represent the rotational temperature associated with the population distribution of the \COB. The red squares represent the rotational temperature calculated from the rotational population distribution of the \CCd state.}
\label{fig:Tscanp}
\end{figure}

In the expanded regime, the emission spectrum is dominated by the CO (\Angstrom) Angstrom bands. Silva et al. \cite{Silva2014} and Du et al. \cite{Du2017} investigated the use of CO Angstrom bands as a thermometric species in CO$_2$ plasmas. They observed that the \COB sate rotational distribution, determined from the analysis of the 0-1 vibrational transition, is in equilibrium with the surrounding gas temperature. Based on these studies, the rotational temperature of the \COB state is used here also as a measure of the gas temperature.

Figure \ref{fig:Tscanp} shows the measured gas temperature determined from \COB and \CCd emission spectra as a function of pressure. The measurements are performed between 60 and 1000 mbar in the middle of the resonator (+20 mm from the resonator bottom, radially centered r=0). The measurement of the C$_2$ Swan band are carried out at constant CO$_2$ flow of 10 L/min and microwave power of 900~W. The emission of the CO molecule is investigated for two conditions: \COO flow 5 L/min and microwave power of 900 W and \COO flow of 10 L/min and microwave power 1500 W. 
In the expanded regime the gas temperature is about 2400 K $\pm$ 200 K at 60 mbar and increases with pressure up to 2800~K $\pm$ 280~K. 

Right before the transition from expanded to contracted regime, the measured gas temperature is 2800~K $\pm$ 280~K at a pressure of 125 mbar. Such measurement is consistent with the values measured by means of Raman scattering at the center of a similar plasma torch by van den Bekerom et. al. \cite{Bekerom2018}. 
In the pressure range that operates in contracted regime (from \ca 120~mbar up to 1000~mbar) similar gas temperatures values have been already measured by Babou et al. \cite{Babou2008}, Spencer et al. \cite{Spencer2012}, Mitsingas et al. \cite{Mitsingas2016}, Bongers et al. \cite{Bongers2017} and Groen et al. \cite{Groen2019}. The vibrational temperature of the \CCd state, in this study, is repeatedly observed to be in equilibrium with the gas temperature suggesting that the heavy particle in the plasma are in equilibrium, when the plasma is contracted. A vibrational temperature in equilibrium with the gas temperature is also measured by Babou et al. \cite{Babou2008} while Spencer et al. \cite{Spencer2012} and Mitsingas et al. measure an  higher vibrational temperature 7700~K, Bongers et al. \cite{Bongers2017} measured an even higher vibrational temperature 9000~K. The reason for such differences in the vibrational can be related to the low sensitivity of the C$_2$ Swan band $\Delta \nu$=0 transition to the vibrational temperature as discussed in Carbone et al. \cite{C2paper}, but also the need of higher resolution spectrometer \cite{C2paper}. To summarize the observation: the plasma transition from expanded to contracted regime is abrupt and the plasma appears to be stable also close to the contraction pressure (no oscillation can be observed). No intermediate temperature can be measured around the transition. The abrupt temperature variation is consistent with the abrupt change of  plasma size and hence the abrupt change in power density.

\subsection{Plasma size}
\label{sec:3.3} 

The axial integrated light emission has been acquired by iCCD imaging using the experimental setup shown in figure \ref{fig:expsetup}~\textit{(b)}. This measurement gives the cross section of the plasma in the radial direction.  The analysis of the plasma emission cannot be analyzed using a single bi-dimensional Gaussian profile but rather using a sum of 2 bi-dimensional Gaussian profiles. However, the separation of the plasma emission into two components has no strong physical basis and their width follow similar behavior changing the power, the \COO flow and the pressure. To reduce the systematic errors in the determination of the plasma region, it has then been defined as the region in which the light intensity is above 15 \% of its maximum. This is sufficient, with a good signal to noise ratio, to identify the region where the C$_2$ molecule, in the contracted regime, and the CO, in the expanded regime, emit light. The gas temperature can be determined from the emission of these molecules, therefore the identified region correspond to the region where the gas temperature is the once discussed in section \ref{sec:3.2}. Since the energy of the MW is coupled to the electrons and transfer by collision to the heavy species \cite{Schulz2012}, the presence of hot gas (i.e. emission from CO or C$_2$) is assumed to be a trace of the presence of electrons, thus of plasma.

\begin{figure}[H]
\centering
\subfloat[][\emph{60 mbar}]{\includegraphics[width=0.48\textwidth]{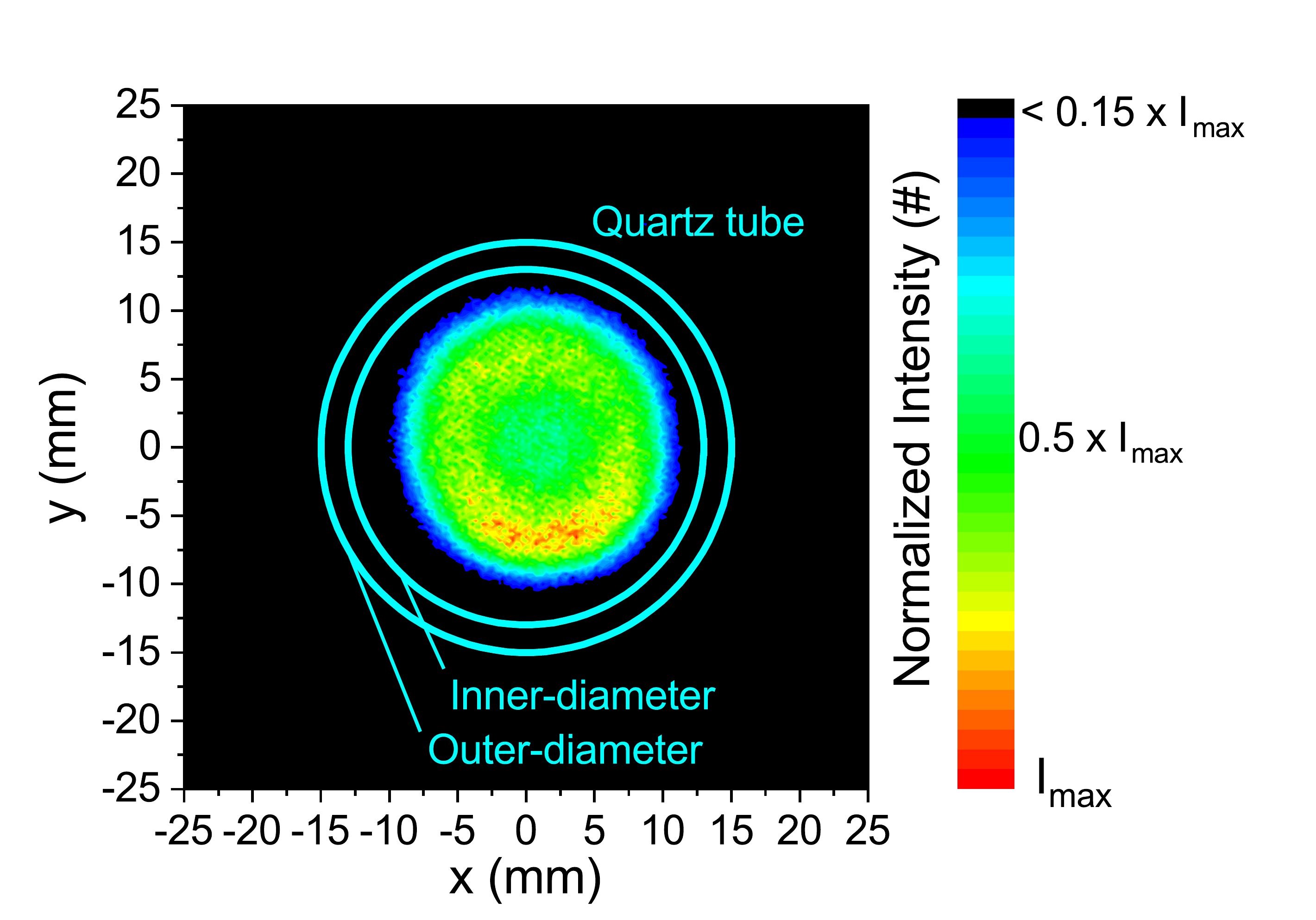}}
\subfloat[][\emph{80 mbar}]{\includegraphics[width=0.48\textwidth]{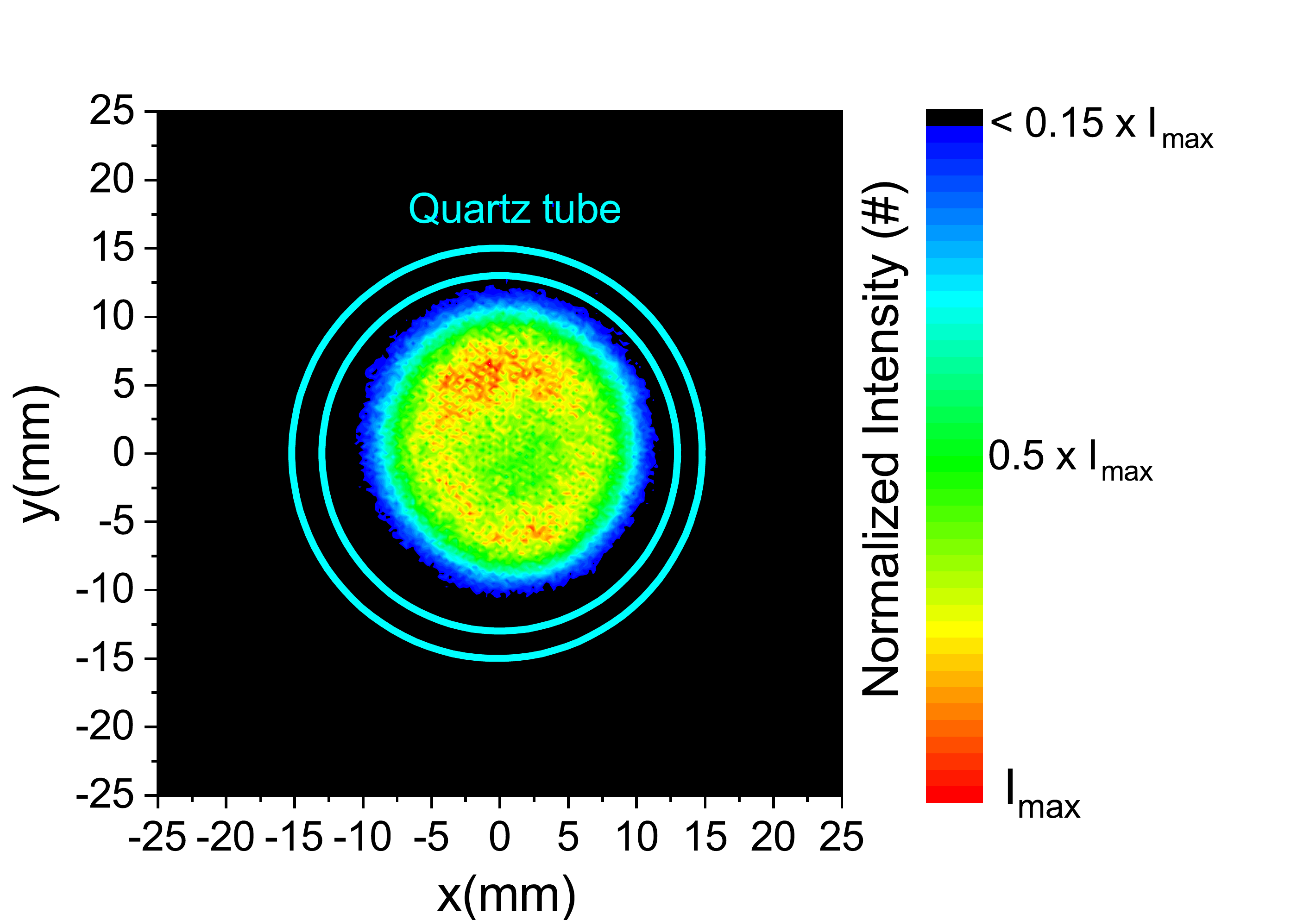}}

\subfloat[][\emph{100 mbar}]{\includegraphics[width=0.48\textwidth]{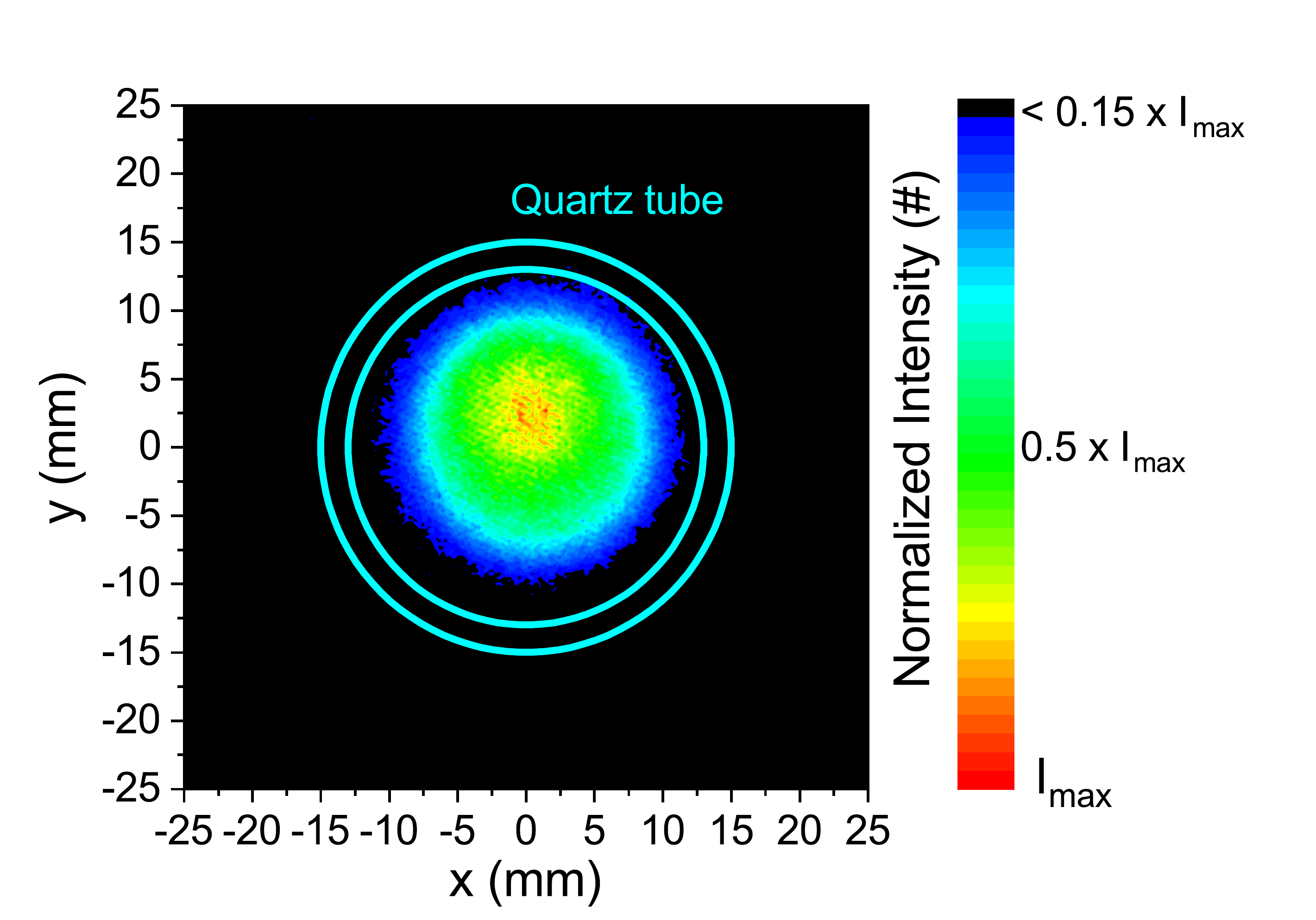}}
\subfloat[][\emph{110 mbar}]{\includegraphics[width=0.48\textwidth]{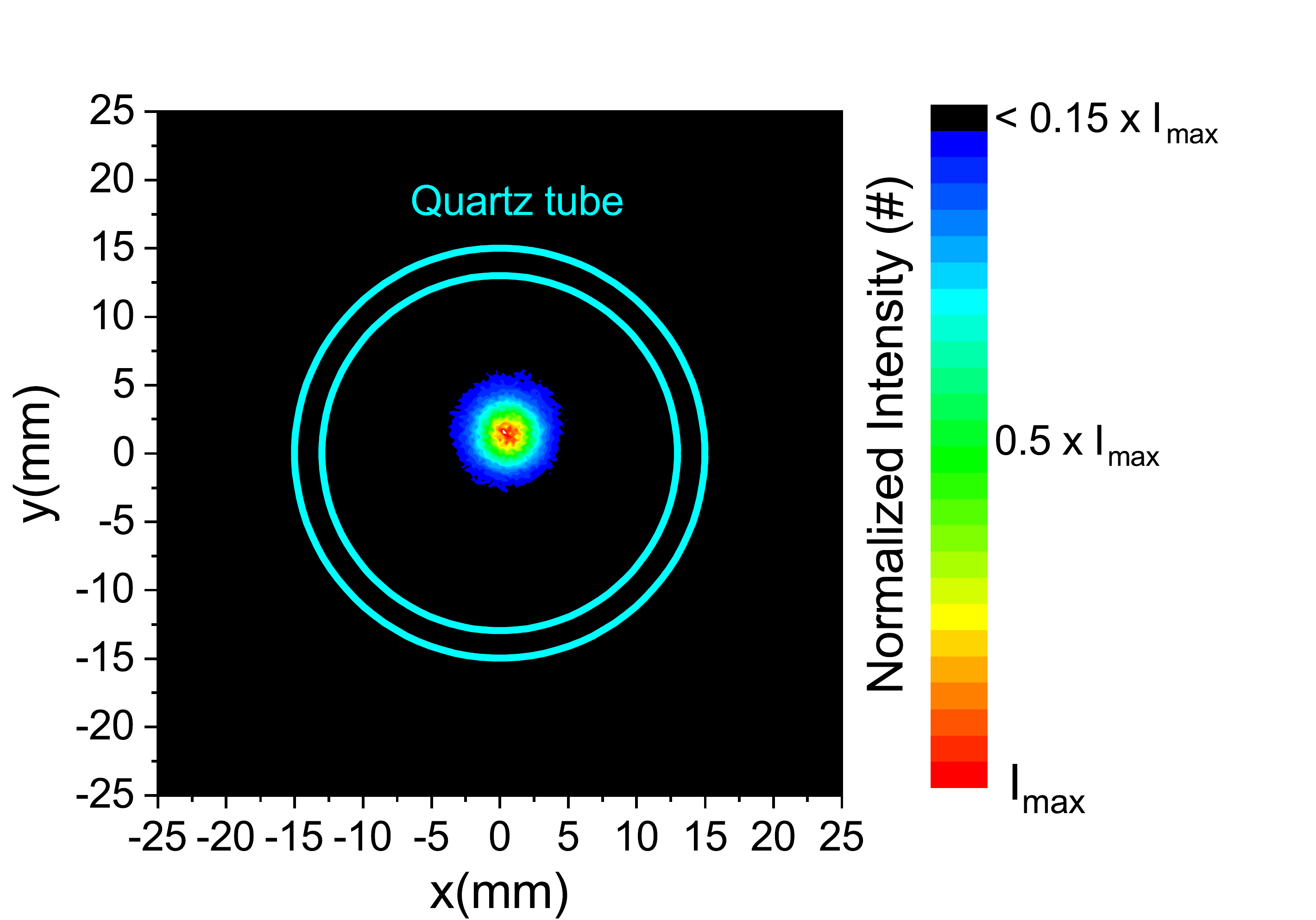}}

\caption{Radial extension of the plasma at 2400 W and 10 L/min at pressure of 60, 80, 100, 110 mbar. The microwave waveguide is positioned at the right side of the picture.}
\label{fig:ICCD-radial}
\end{figure}

Figure \ref{fig:ICCD-radial} shows a measurement of the axially integrated light emission performed with the iCCD camera performed at 10 L/min of \COO injection and a microwave power of 2400 W and pressures range from 60 to 110~mbar. At pressure of 60 and 80 mbar (figure \ref{fig:ICCD-radial}~\textit{(a),(b)}) the plasma emission shows an hollow profile, whereas at pressure of 100~mbar the plasma emission peaks in the center of the quartz tube (figure \ref{fig:ICCD-radial}~\textit{(c)}). The transition depends also on the power coupled to the plasma, at 60 mbar, the plasma emission peaks in the center of the quartz tube for any power below \ca 1500~W. 

\begin{figure}[H]
\centering
\subfloat[][\emph{}]{\includegraphics[width=0.48\textwidth]{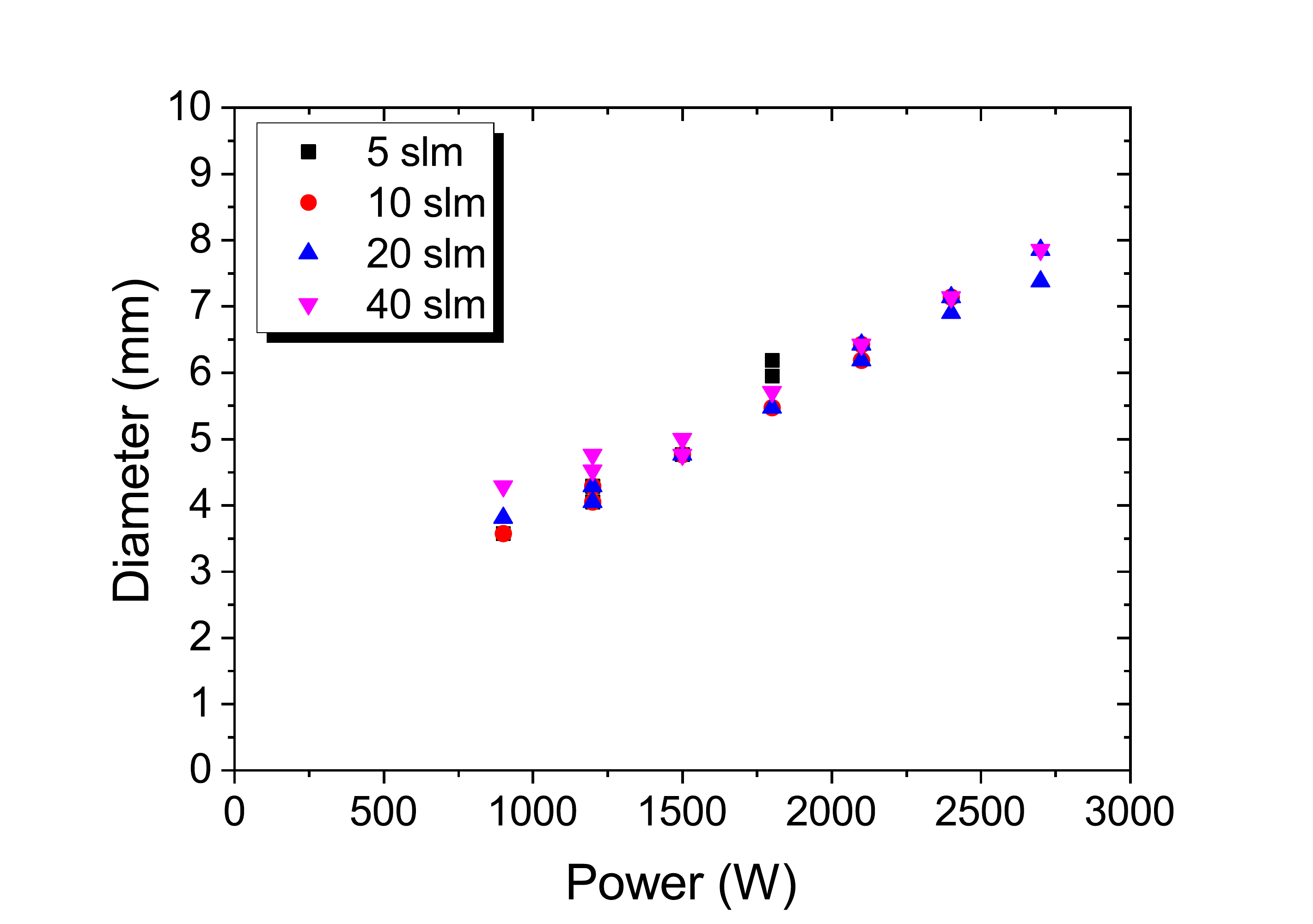}}
\subfloat[][\emph{}]{\includegraphics[width=0.48\textwidth]{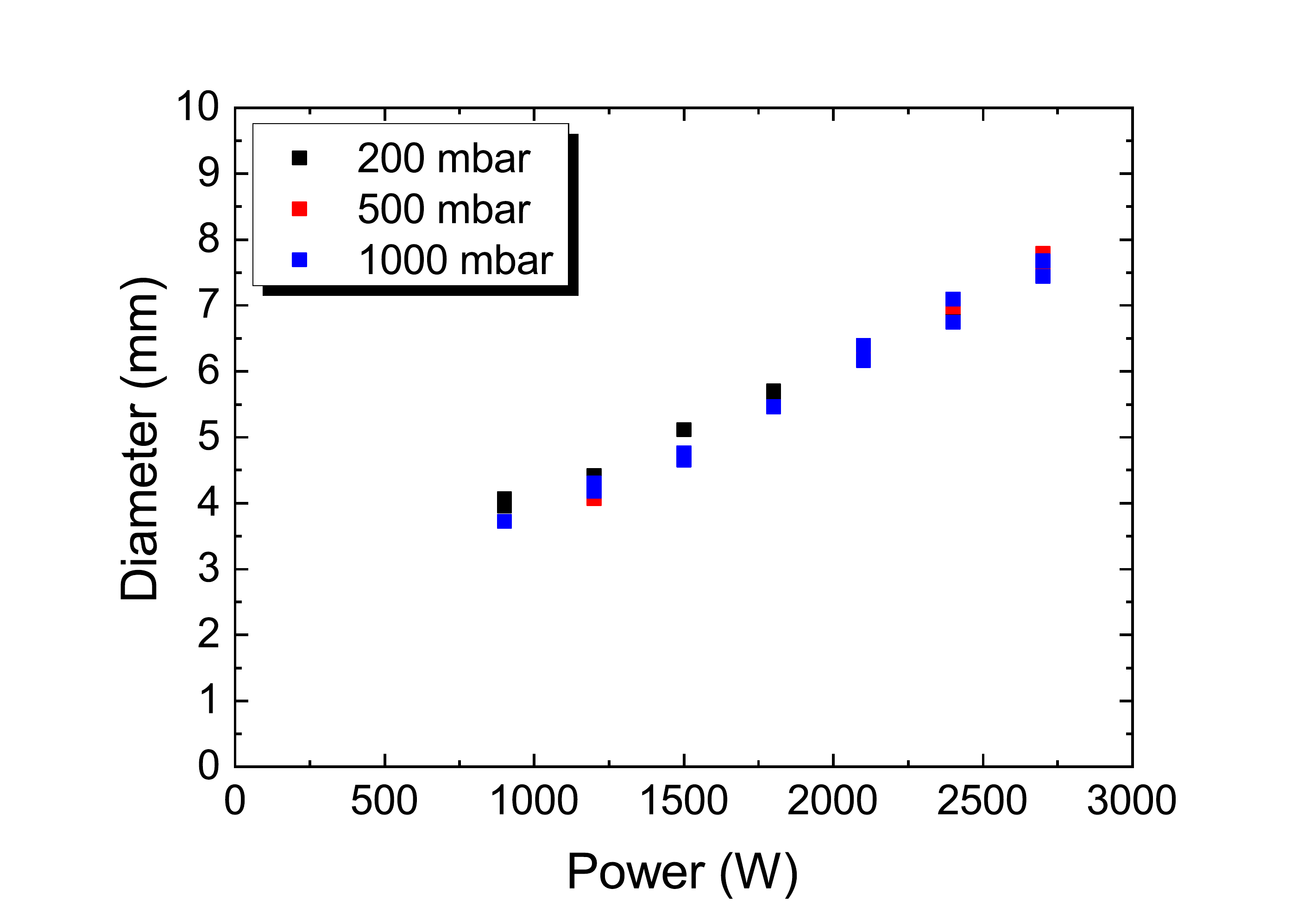}}

\caption{ Figure \textit{(a)} shows the changes in the plasma cross section at constant pressure above 900 mbar and different flows. Figure \textit{(b)} shows the effect of the pressure on the plasma cross section. }
\label{fig:pscandiameter}
\end{figure}

After the transition to a contracted regime (i.e. filamentary plasma) the plasma cross section abruptly and drastically reduces: the plasma diameter changea from of about 20~mm in the expanded regime to values below 10~mm. Figure \ref{fig:pscandiameter} \textit{(a)} shows the changes of plasma diameter with power at different flows and constant pressure above 900 mbar. 
Figure \ref{fig:pscandiameter} \textit{(b)} shows the changes of the plasma diameter with power at pressure of 200, 500 and 900 mbar for a constant flow of 20 L/min. The cross section of the plasma weakly decreases while increasing the pressure. The dominant parameter that defines the plasma cross section, after the plasma contraction takes place, is the microwave power coupled into the plasma.     

\begin{figure}[H]
\centering
\subfloat[][\emph{\\Plasma}]{\includegraphics[height=0.6\textwidth]{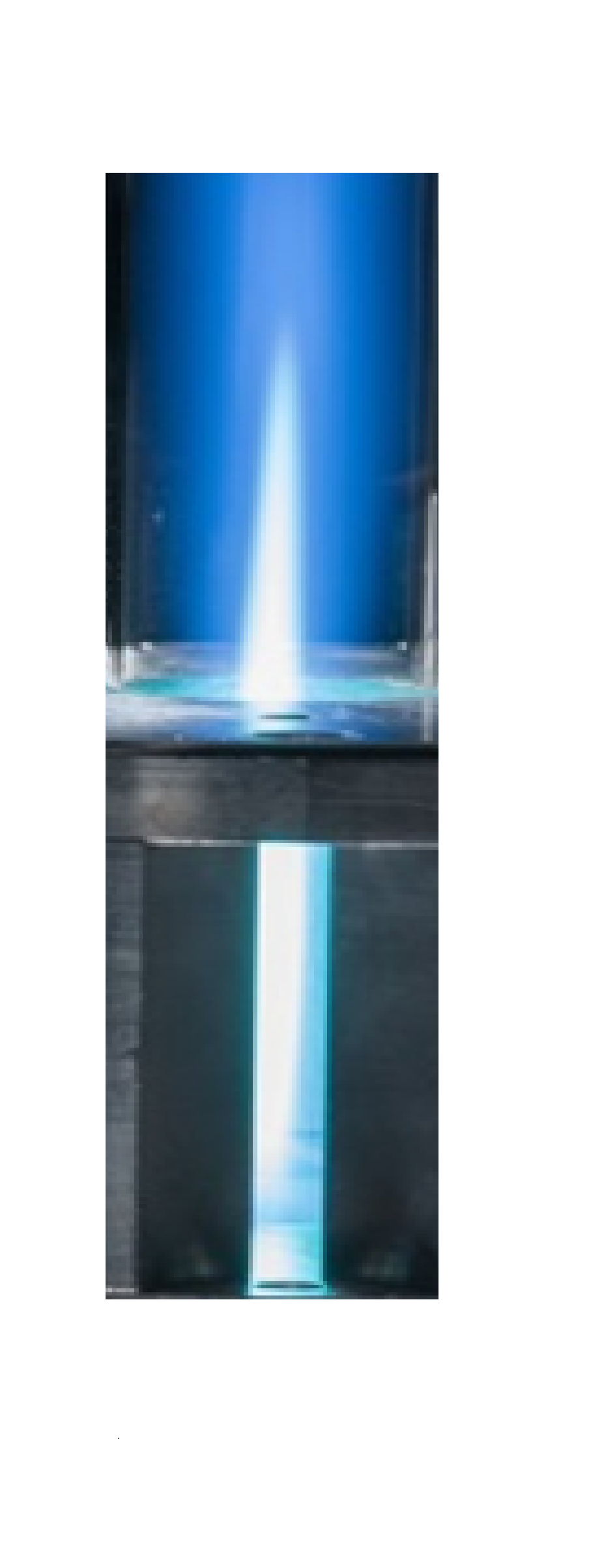}}
\subfloat[][\emph{60 mbar,\\ 2700 W}]{\includegraphics[height=0.6\textwidth]{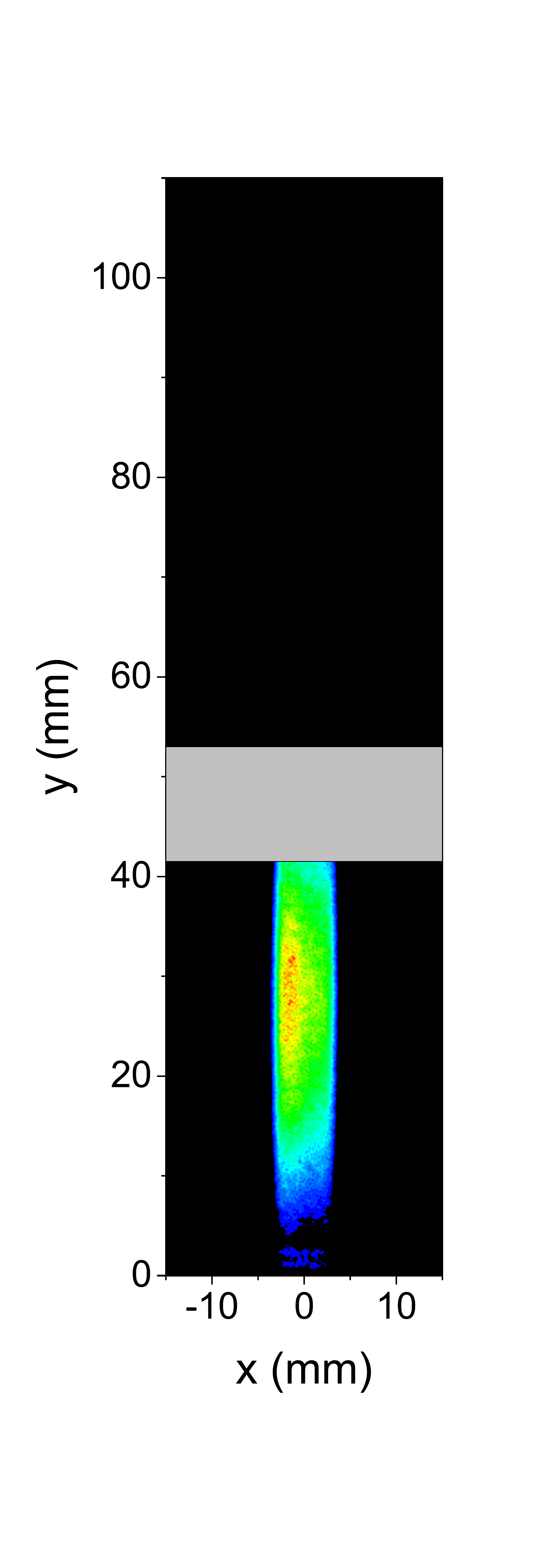}}
\subfloat[][\emph{200 mbar,\\ 2700 W}]{\includegraphics[height=0.6\textwidth]{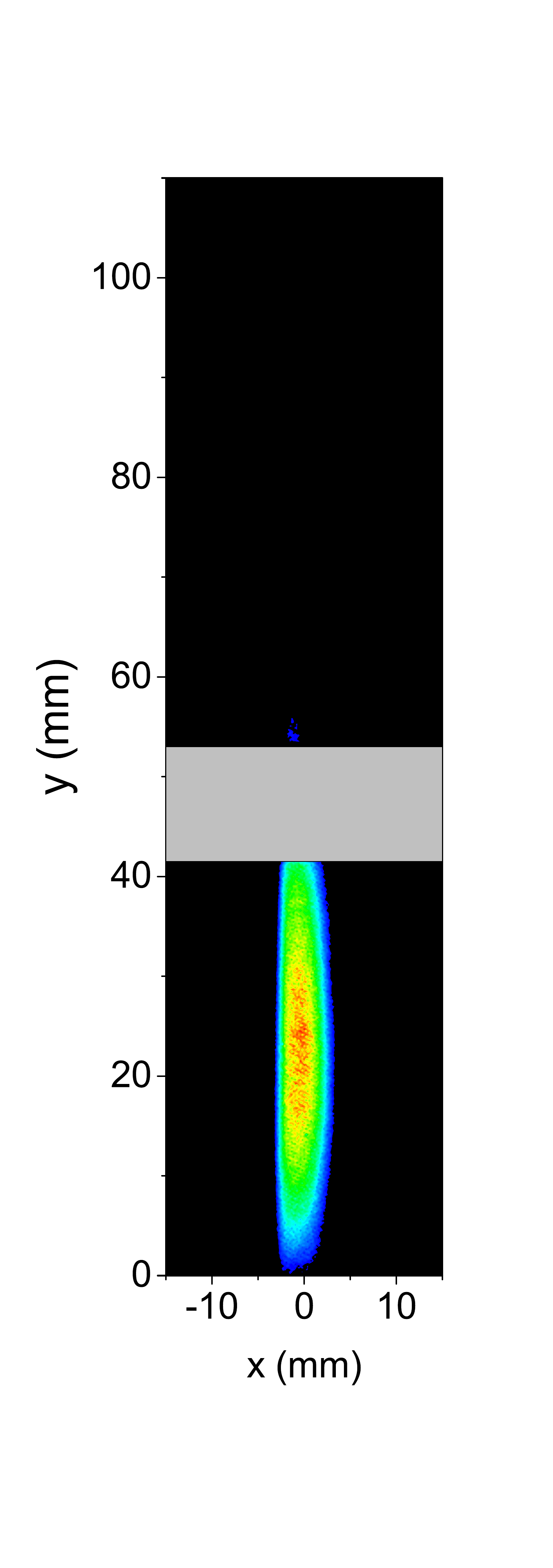}}
\subfloat[][\emph{920 mbar,\\ 900 W}]{\includegraphics[height=0.6\textwidth]{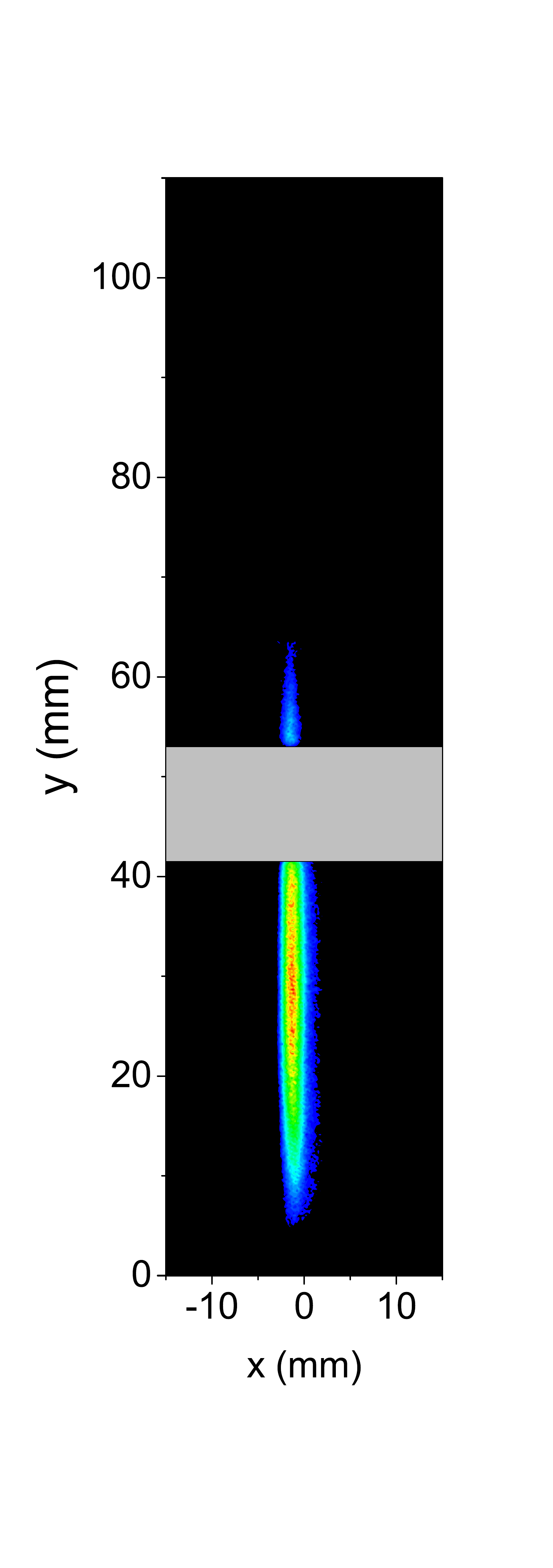}}
\subfloat[][\emph{920 mbar,\\ 2700 W}]{\includegraphics[height=0.6\textwidth]{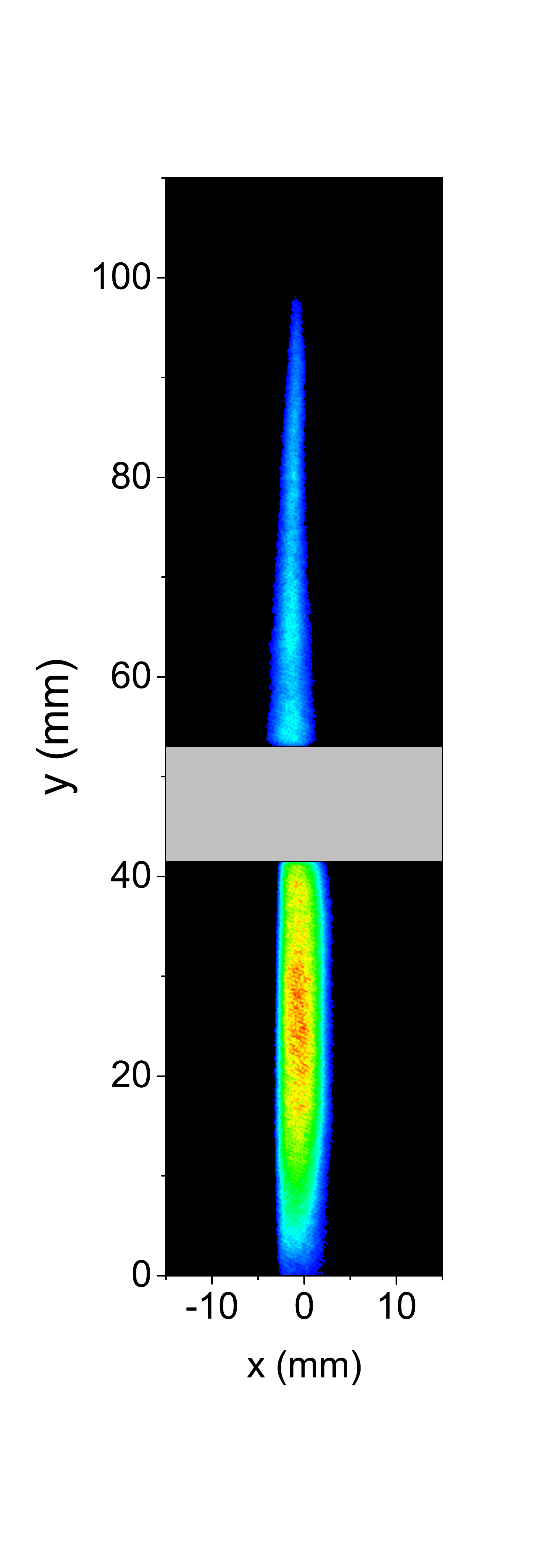}}

\caption{Figure \textit{(a)} shows a photo of the plasma burning in the microwave cavity. Axial ICCD images of the plasma at 10 L/min of \COO flow, pressure of 60 mbar and microwave power of 2700 W \textit{(b)}, 200 mbar and 2700 W \textit{(c)}, quasi-atmospheric pressure 900 W \textit{(d)} and 2700 W \textit{(e)}.}
\label{fig:ICCD-axial}
\end{figure}

The plasma extension in the axial direction is recorded with the ICCD camera mounted on the side of the microwave resonator. In the resonator the light emission is limited by the slit size (5.5 mm). 
Figure \ref{fig:ICCD-axial} \textit{(a)} shows an optical picture of the plasma taken from the side in which the masking due the waveguide wall is visible. Similarly to the cross section, the plasma extension in the axial direction (plasma length) was determined as the region where the emission is above 15 \% of its maximum. In figure  \ref{fig:ICCD-axial}, some typical radial integrated plasma optical emission profile are shown for different pairs of pressure and power. The plasma extends in the effluent at atmospheric pressure (see pictures \textit{(d), (e)}) and at low power does not fill the bottom of the resonator \textit{(c)}. Reducing the pressure the plasma extends much less in the effluent, as can be seen while comparing figures \ref{fig:ICCD-axial} \textit{(c)} and \textit{(e)}. Figure \ref{fig:ICCD-axial} \textit{(b)} shows, on the other hand, that in the expanded regime the plasma fills the resonator without extending above it.

\begin{figure}[H]
\centering
\subfloat[][\emph{}]{\includegraphics[width=0.48\textwidth]{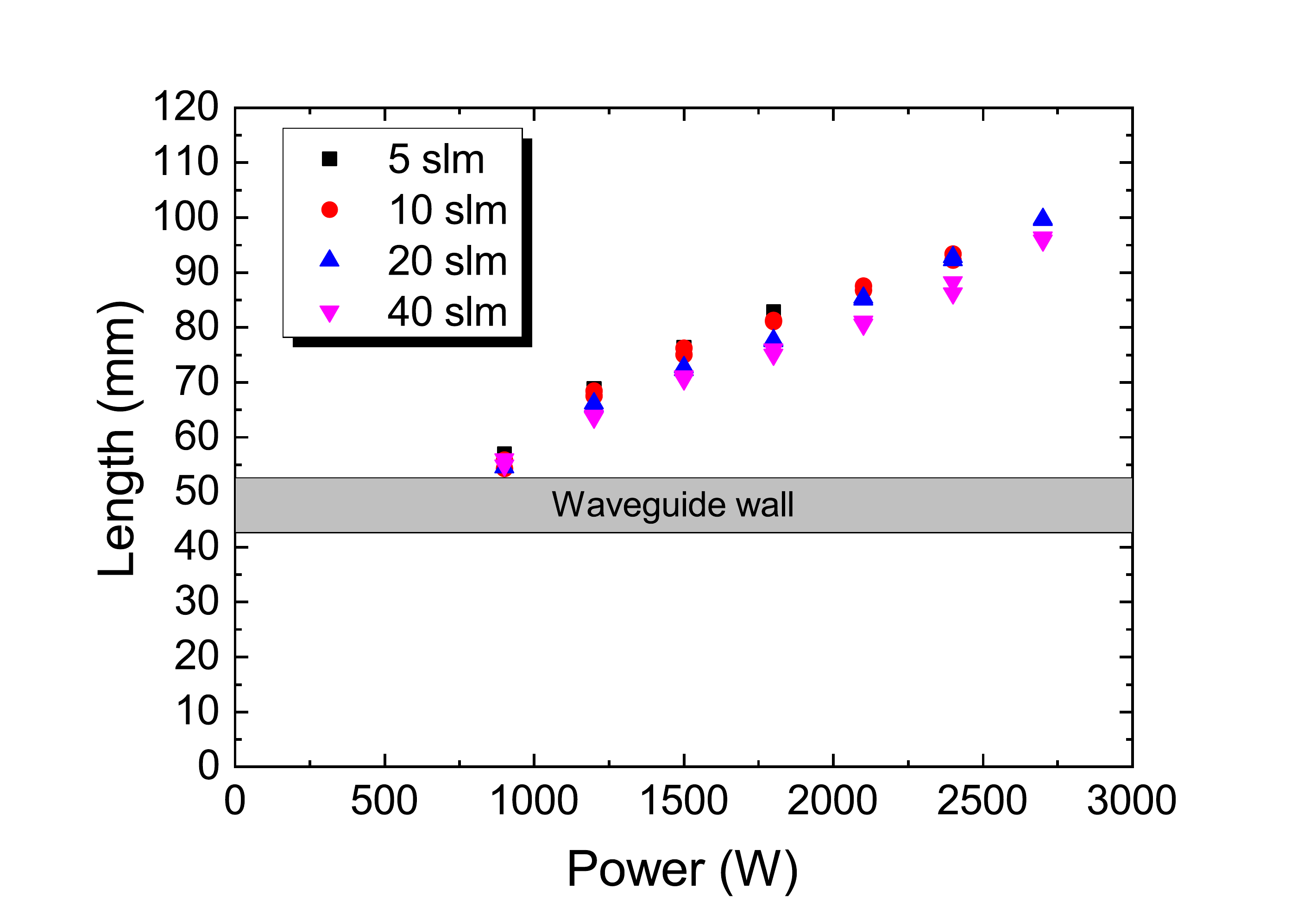}}
\subfloat[][\emph{}]{\includegraphics[width=0.48\textwidth]{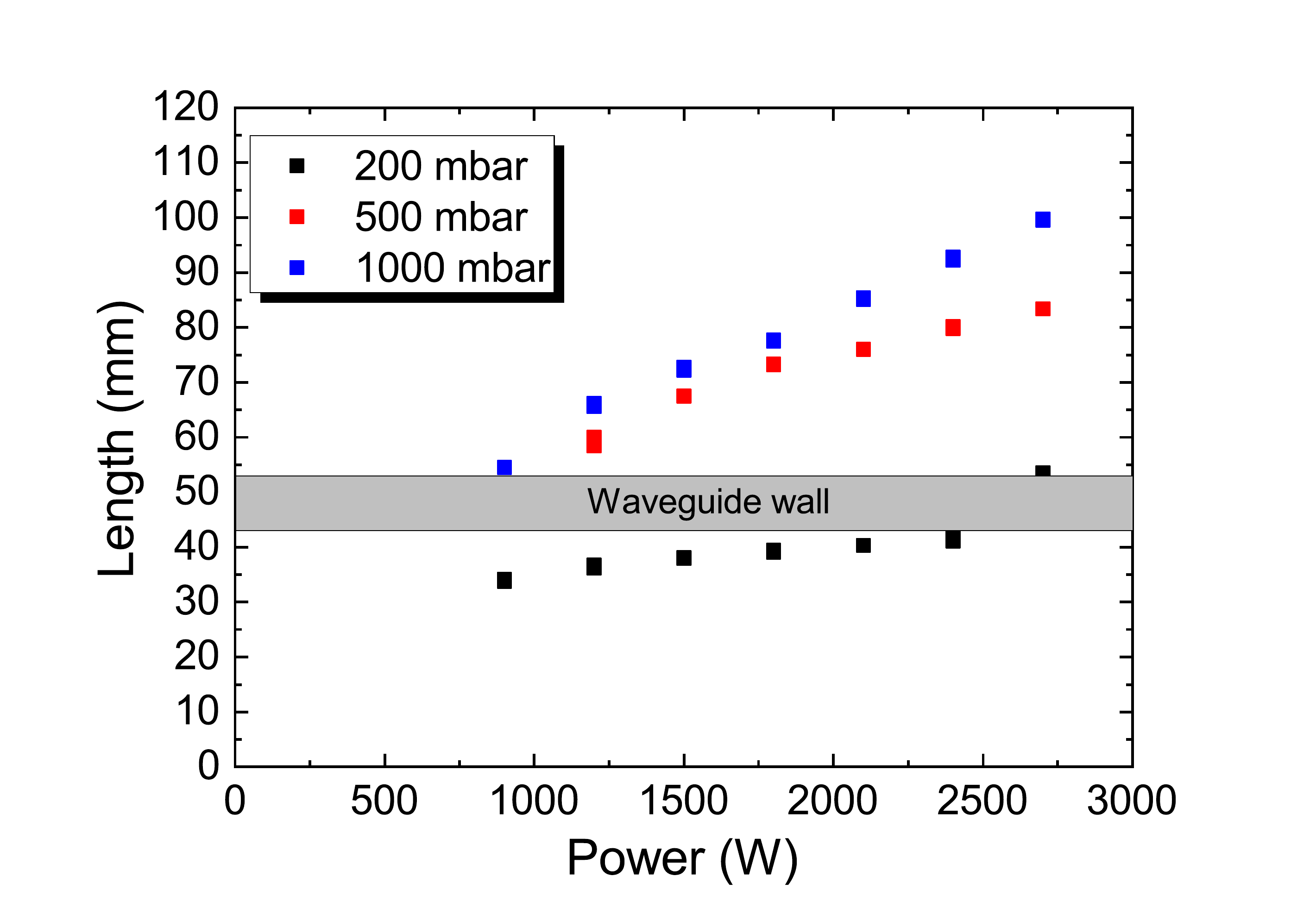}}

\caption{Figure \textit{(a)} shows the changes of plasma length as function of power for several \COO flows at a constant pressure above 900~mbar. Figure \textit{(b)} shows the plasma axial extension at different pressures and a constant flow of 20 L/min. In both figures the horizontal dashed line indicates the end of the resonator.}
\label{fig:pscanaxial}
\end{figure}

A study of the plasma length is performed and the results are summarized in figure \ref{fig:pscanaxial}. The plasma length as function of power and for several flows is reported in figure \ref{fig:pscanaxial} \fa for near atmospheric pressure conditions (i.e. about 900~mbar). The flow only weakly influences the plasma length while the plasma length significantly increases with power.  Figure \ref{fig:pscanaxial} \textit{(b)} shows the effect of the power at different pressures and constant flow 20 L/min. The plasma length strongly increases with pressure. In the investigated power range at 200 mbar the axial elongation is mostly not visible because the microwave waveguide edge covers the variation of plasma emission. At pressure below 200 mbar the plasma mostly expands radially, increasing its cross section. At higher pressure the plasma mostly elongates axially (but the radial extension is comparable to the ones observed at 200~mbar). The total plasma volume, calculated as the volume of the cylinder having the measured cross section and length, at high pressure depends (mostly) on the plasma radius. At pressure below 100~mbar the plasma appears to fill the quartz tube, although some small variation of the diameter can be observed (18-22~mm). The diameter variation at this pressure is the only responsible of the volume changes.

The SEI used in section \ref{sec:3.1} is calculated on top of the total \COO flow and microwave power coupled into the plasma. The plasma emission can also be used as a measure of the plasma volume and define a region of gas swirling around a hot plasma core. In that case, the local SEI for the molecules entering the plasma region is significantly higher than the global SEI based on the total gas flow entering the quartz tube. For the present setup, the typical local specific energy input calculated, as suggested by van den Bekerom \cite{Bekerom2018}, as the global SEI divided by the fraction of volume occupied by the plasma is always (much) above 10 eV/molecule in the contracted regime. With such high local SEI, and gas temperatures as reported in the previous section, one can expect that energy efficiencies will be quite low and this is what is indeed observed experimentally (cf. section \ref{sec:3.1}). 

\label{sec:3.0} 

\section{Discussion}

\subsection{Thermal equilibrium consideration}
\label{sec:thermo}
\begin{figure}[H]
\centering
\includegraphics[width=0.55\textwidth]{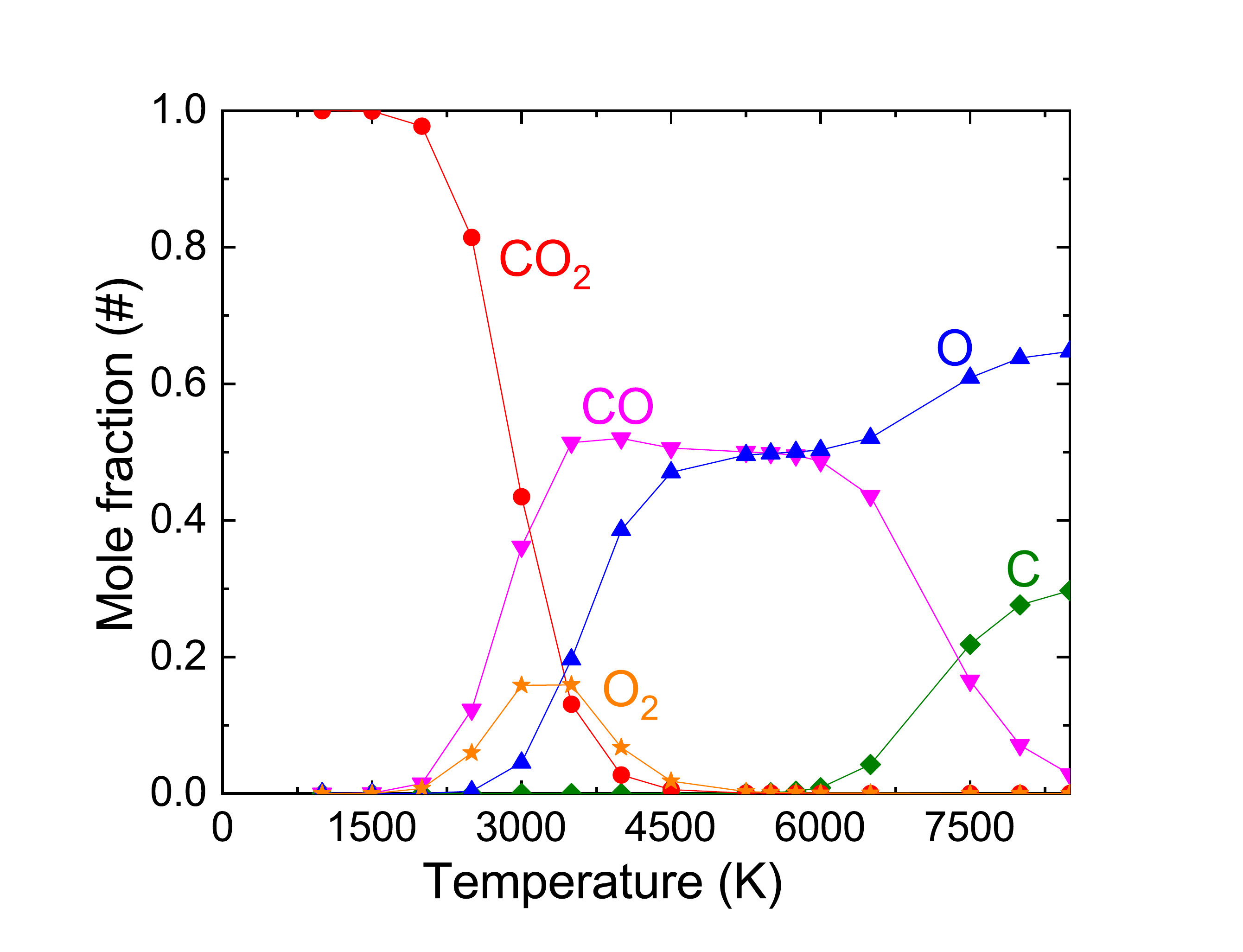}
\caption{The mole fraction of \COO (red), CO (magenta), O$_2$ (orange), O (blue) and C (green) expected by heating a mole of \COO at atmospheric pressure to a given temperature. The dots correspond to the temperature at which the composition is calculated.}
\label{fig:species}
\end{figure}

In addition to electron driven processes, the contribution of thermal conversion of \COO into CO needs to be considered when a gas temperature above 2000~K is measured. To evaluate the degree of \COO dissociation in the core of the plasma, thermal equilibrium calculations have been performed using the program CEA \cite{CEA}. It can calculate the chemical composition of a gas (or a gas mixture) at fix pressure and temperature. To understand the variation of chemical composition as function of the temperature in the investigated pressure range (60~mbar - quasi atmopspheric pressure), the thermal equilibrium has been computed at several combinations of pressure and temperature. Figure \ref{fig:species} shows the molar fractions of \COO, CO, O$_2$, O and C as a function of gas temperature while heating up a pure \COO gas. The molar fraction of \COO decreases rapidly above temperatures of about 3000~K with formation of CO and O$_2$. Above this temperature O$_2$ dissociate and the gas becomes predominantly a CO+O mixtures. At temperatures above 6000~K, CO starts dissociating and the gas becomes a mixtures of C and O atoms. It is then no coincidence to see that C$_2$ Swan bands appearance is correlated with gas temperatures of about 6000~K, as recombination processes of C atoms lead to the formation of the C$_2$ Swan bands in CO$_2$ plasmas \cite{C2paper}. 

\begin{figure}[H]
\centering

\subfloat[][\emph{}]{\includegraphics[width=0.48\textwidth]{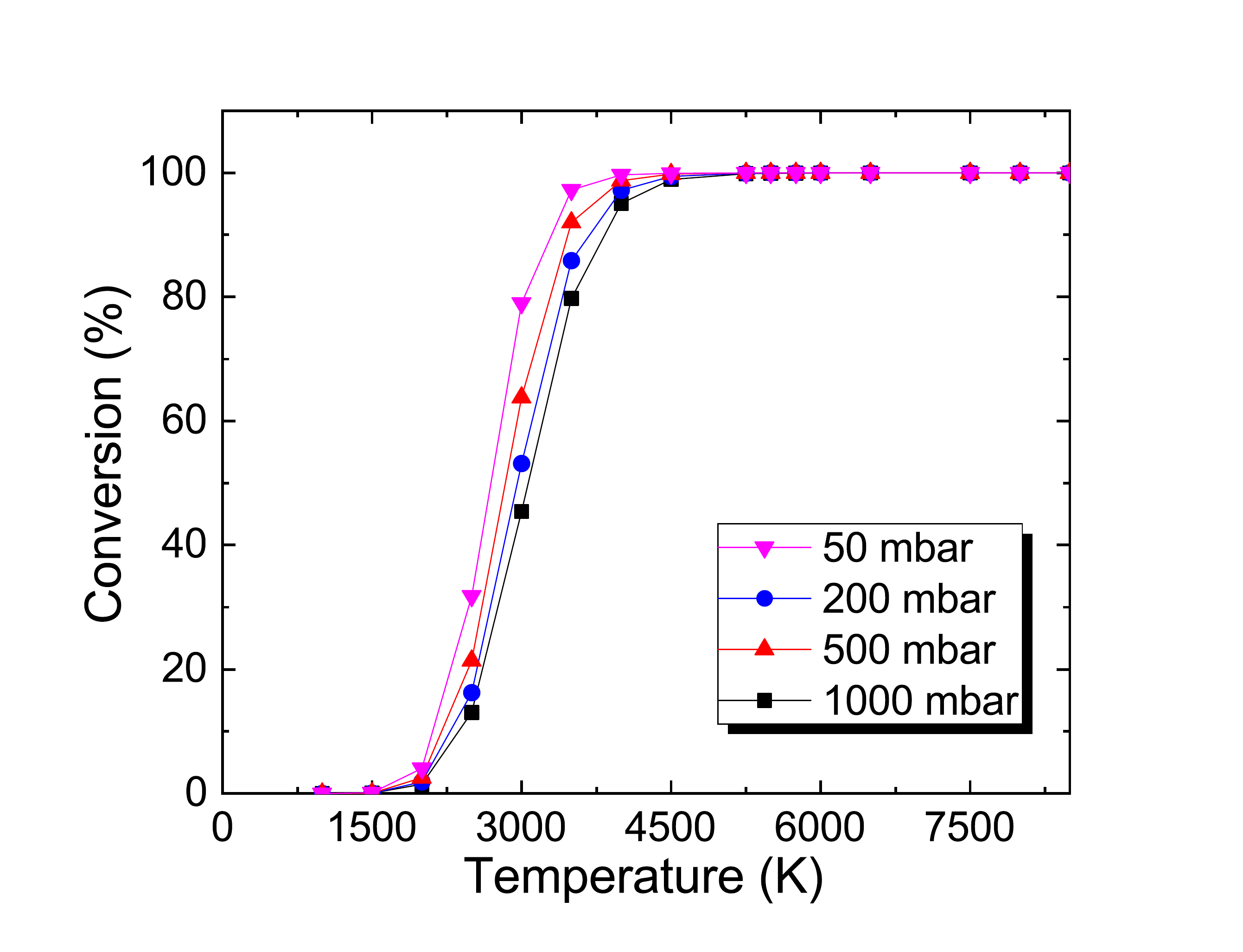}}
\subfloat[][\emph{}]{\includegraphics[width=0.48\textwidth]{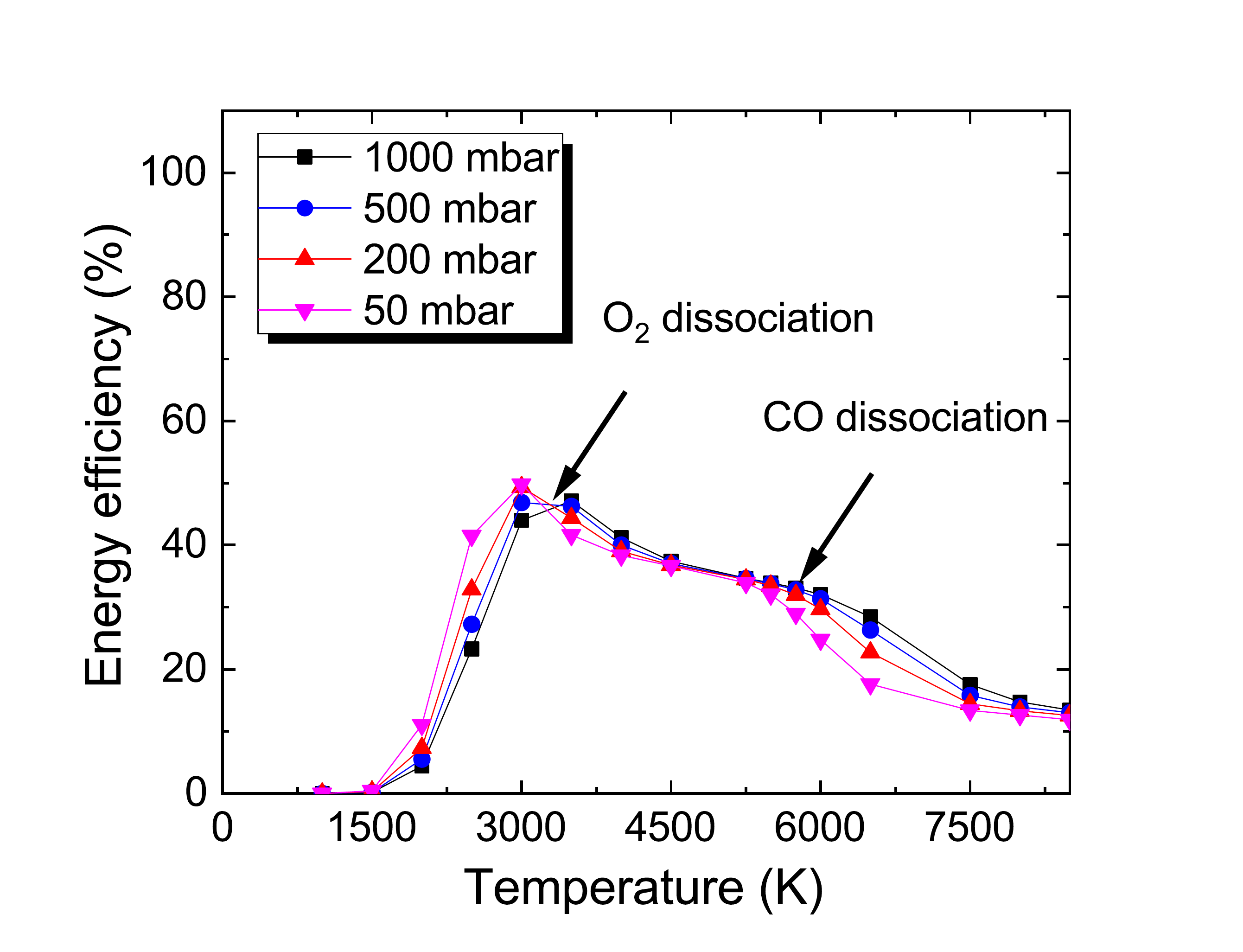}}

\caption{Plot \textit{(a)} shows the \COO dissociation fraction expected at a given pressure as function of the gas temperature, under the assumption of ideal quenching (no \COO losses) and recombination of carbon atoms into CO. Plot \textit{(b)} shows the expected energy efficiency at a given pressure as function of the temperature. 
}
\label{fig:Thermal}
\end{figure}

Figure \ref{fig:Thermal} \textit{(a)} shows the degree of conversion of \COO into  CO as function of the gas temperature for different given pressures. Note that the conversion is calculated under the assumption that all the carbon atoms (that are formed at T$_{gas}>$5500 K) recombine into CO. The pressure has only a weak effect on the conversion. For instance from 50 to 1000~bar, the gas temperature required to convert 50\% of the CO$_2$ into CO increases from 2700 to 3050~K.

The calculated conversion refers to the \COO dissociation fraction. The energy spend to heat the \COO and used in the conversion of \COO into CO can be expressed as \ref{eq:Enthalpy}: 

\begin{equation}
Q = \int_{T_0}^{T_{fin}} \left(\frac{\partial H}{\partial T}\right)_p \: dT  = H(T_{fin}) - H(T_0) 
\label{eq:Enthalpy}
\end{equation}
where T$_0$ and T$_{fin}$ represent the initial temperature (298 K) and the final temperature, H is the enthalpy of the gas mixture at a given temperature. The enthalpy  of the mixture is a result of the CEA calculation, from which the energy per molecule required to obtain a given temperature can be calculated:

\begin{equation}
SEI = \frac{Q [J/g] \cdot 44.07 [g/mol]}{e [J/eV] N_A [molecule/mol]}
\label{eq:SEI-CEA}
\end{equation}

where Q is given in J/g, 44.07 is the molar mass of \COO, $N_A$ the Avogadro number and $e$ the conversion factor between J and eV, hence the units of the SEI are eV/molecule. The latter is used to calculate the energy efficiency shown in figure \ref{fig:Thermal} \textit{(b)}. The energy efficiency of thermal dissociation of \COO into CO is shown in figure \ref{fig:Thermal} \textit{(b)} for the same parameters.
The energy efficiency of thermal dissociation for producing CO peaks at about 3000~K and degrades at higher temperatures where energy is then spent not only for warming up the gas (i.e. via its heat capacity) but also for dissociating O$_2$ and CO (see figure \ref{fig:species}).

With the help of the thermal equilibrium calculation it is possible to investigate the importance of thermal dissociation in the measured conversion.

The gas temperature measured at 60 mbar in the microwave resonator is \ca 2500 K, where a \COO conversion of 30 \% is expected by thermal dissociation only. The maximum measured conversion rate at 60~mbar for the present setup is however only 25\% (see figure \ref{fig:pscan}). The lower measured conversion can be related to the gas temperature gradient, since not all the gas is heated to the measured temperature, but part of it is colder, thus a lower conversion can be expected. However to observe in an experiment the dissociation fractions given by figure \ref{fig:Thermal} \fa, the mixture composition needs to be quenched rapidly to avoid recombination of CO via:

\begin{equation}
CO + O + M \rightarrow CO_2 + M.
\label{eq:CORec}
\end{equation}

In order to minimize the losses the gas needs to be cooled down very rapidly (faster then 10$^6$ K/s \cite{Fridman2008}) while exiting the plasma volume, since the process described in equation \ref{eq:CORec} is temperature dependent, and the reaction rates is proportional to $e^{-\frac{1500}{T_{gas}}}$ \cite{Tsang1986}. This process typically happens in the plasma itself, in the gas layer that surrounds the plasma and in the effluent as evidence by its chemiluminescence continuum reported in section \ref{sec:3.2}. 
The increase of converted \COO with pressure that can be observed in figure \ref{fig:pscan} \textit{(a)} can be correlated with the observed increase of gas temperature that is measured with power for a non contracteed plasma. As the pressure increases the gas temperature increases with a maximum of 2800~K measured experimentally right before the transition to a contracted regime. Assuming a thermally driven dissociation process, thermal calculations do predict a increasing energy efficiency with gas temperature as seen in figure \ref{fig:Thermal} \fb. On the other hand the pressure has only a limited effect on the absolute value of the energy efficiency for thermal dissociation, the gas temperature has the stronger influence on energy efficiency. The observed increase in conversion with power has to be also compared with the plasma size and low overall conversion rate. Moreover a pressure increase at gas temperature below 3000~K (see figure \ref{fig:Thermal} \fb) would actually reduce (slightly) the energy efficiency, while an increase of energy efficiency is observed here (see figure \ref{fig:pscan} \textit{(a)}), indicating that the gas temperature effect would be the driving force of conversion.  The highest energy efficiency is measured at \ca 120~mbar (at the plasma contraction) in good agreement with previously reported trends by Kurchatov institute, but with lower overall energy efficiencies \cite{Fridman2008}. The measured energy efficiencies are maximum 30 \% which is much lower than the 80\% reported by Butylkin et al. \cite{Butylkin1981}, where vibrational ladder climbing was identified as mechanism for \COO dissociation. The overall lower energy efficiency can be explained, by the dissociation mechanism, since in the present setup thermal dissociation is the dominant dissociation mechanism, thus the maximum energy efficiency is limited by the thermal dissociation efficiency. The known differences between the two setups are: the inlet gas velocity (\ca $10^4$ cm/s) reported by Butylkin et al. \cite{Butylkin1981} is higher than the one expected in the present setup (\ca $10^2-10^3$ cm/s), the microwave frequency 2.4 GHz and the wave-guide components size. However if and how these geometrical differences influence the dissociation mechanism of \COO is not understood.  

The plasma contraction increases significantly the power density, that drives an abrupt increase of the gas temperature which reaches a value of 6000~K~$\pm$~500~K and remains constant at every flow, power and pressure investigated. Thermal calculations predict at such high temperature that energy efficiency should increase with increasing pressure. The opposite trend is however found experimentally (see figure \ref{fig:pscan}). Losses of CO by recombination via reaction \ref{eq:CORec} is a pressure depend process (via the third body collisional term in the rate coefficient). Even if locally the conversion of \COO into CO can be high, the recombination process reduces the CO molar fraction measured downstream of the plasma reactor. At low gas flow rates (i.e. high specific energy input as shown in figure \ref{fig:MS}), one can expect a slower cooling rate meaning that a larger fraction of CO is lost. A decrease of conversion and energy efficiency are indeed observed at low gas flows ($<$ 10 L/min) which is more pronounced at high pressure. 
Finally another effect should be taken in account, the plasma elongation in the effluent is correlated with an increase of microwave radiation field into the room (it has been measured using a microwave-power meter). This indicates the generation of a surface-wave: in the absence of a Faraday shield around the quartz tube, a surface wave discharge does not absorb all the power from the applicator and radiates part of its energy as an antenna \cite{Moisan2018}. Such effect would effectively reduce the power coupled to the plasma. If a lower power is coupled to the plasma a the fraction of \COO converted reduces (as seen in figure \ref{fig:MS} ) and the energy efficiency underestimated by overestimating the SEI. 

In the contracted regime the plasma however occupies only a small part of the tube and hydrodynamic effect between the cold swirling gas flow and hot plasma core should be additionally considered (see below for more discussion).  

\subsection{Flow dynamics considerations}

\begin{figure}[H]
\centering
\includegraphics[width=0.55\textwidth]{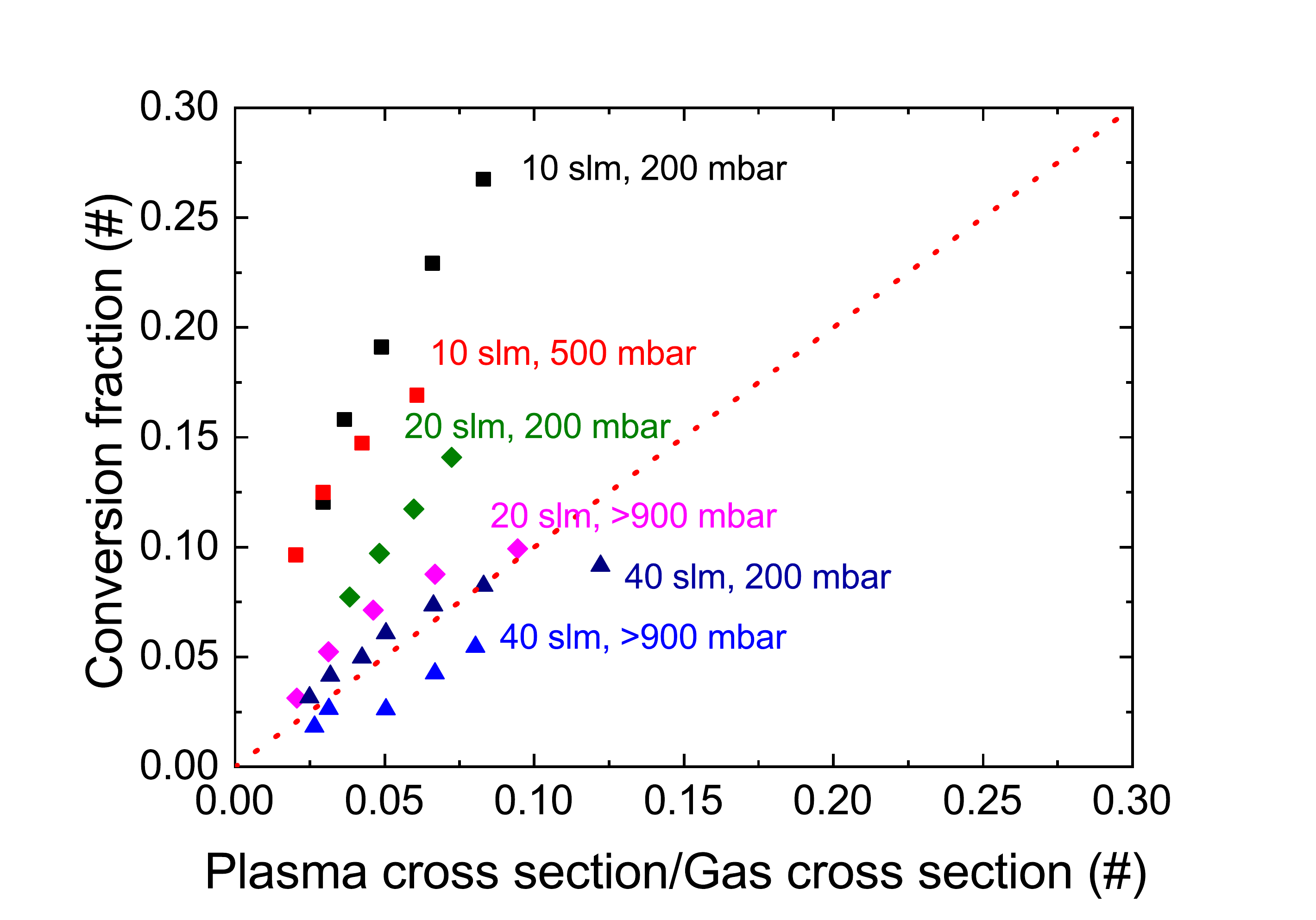}
\caption{The \COO conversion as function of the ratio between plasma volume and gas volume. The red dashed line represent the equivalence between fraction of \COO converted and the ratio of the plasma cross section and gas cross section.}
\label{fig:volumeratio}
\end{figure}

Following the assumption of local chemical equilibrium, the temperature profile in the quartz tube defines the region where CO$_2$ dissociation takes place. Babou at al. \cite{Babou2008} showed that temperature gradients at the edge a N$_2$-CO$_2$ microwave plasma are relatively steep. Similar observation was done by van den Bekerom et al. \cite{Bekerom2018} in the expanded regime. 
If we assume, based on these observation, that the temperature gradient are strong the non-emitting region cannot contribute significantly to the conversion. The conversion observed can be related to mixing between the cold gas and the hot gas driven by the temperature gradient. The amount of \COO that flows from the cold region into the (hot) plasma should then be proportional to the surface (or volume) of the hot region itself.  
Figure \ref{fig:volumeratio} shows the \COO conversion fraction as function of the ratio between the plasma cross section and the gas cross section in several experimental conditions. The plasma cross section is the once discussed in section \ref{sec:3.3}, the gas cross section is the cross section of the quartz tube from which the plasma cross section has been subtracted. The ratio between plasma and gas cross sections does not show any direct correlation with the conversion. only at high flow a correlation can be found, which may be purely coincidental. It should be also note that at 100~slm the plasma does not have any longer a cigar shape, but it is twisted by the strong swirl. These observations indicate that gas mixing dynamics at the edge of the plasma probably represent the key effect to understand the \COO conversion in the high pressure regime. 
 
\begin{figure}[H]
\centering

\subfloat[][\emph{200 mbar}]{\includegraphics[width=0.48\textwidth]{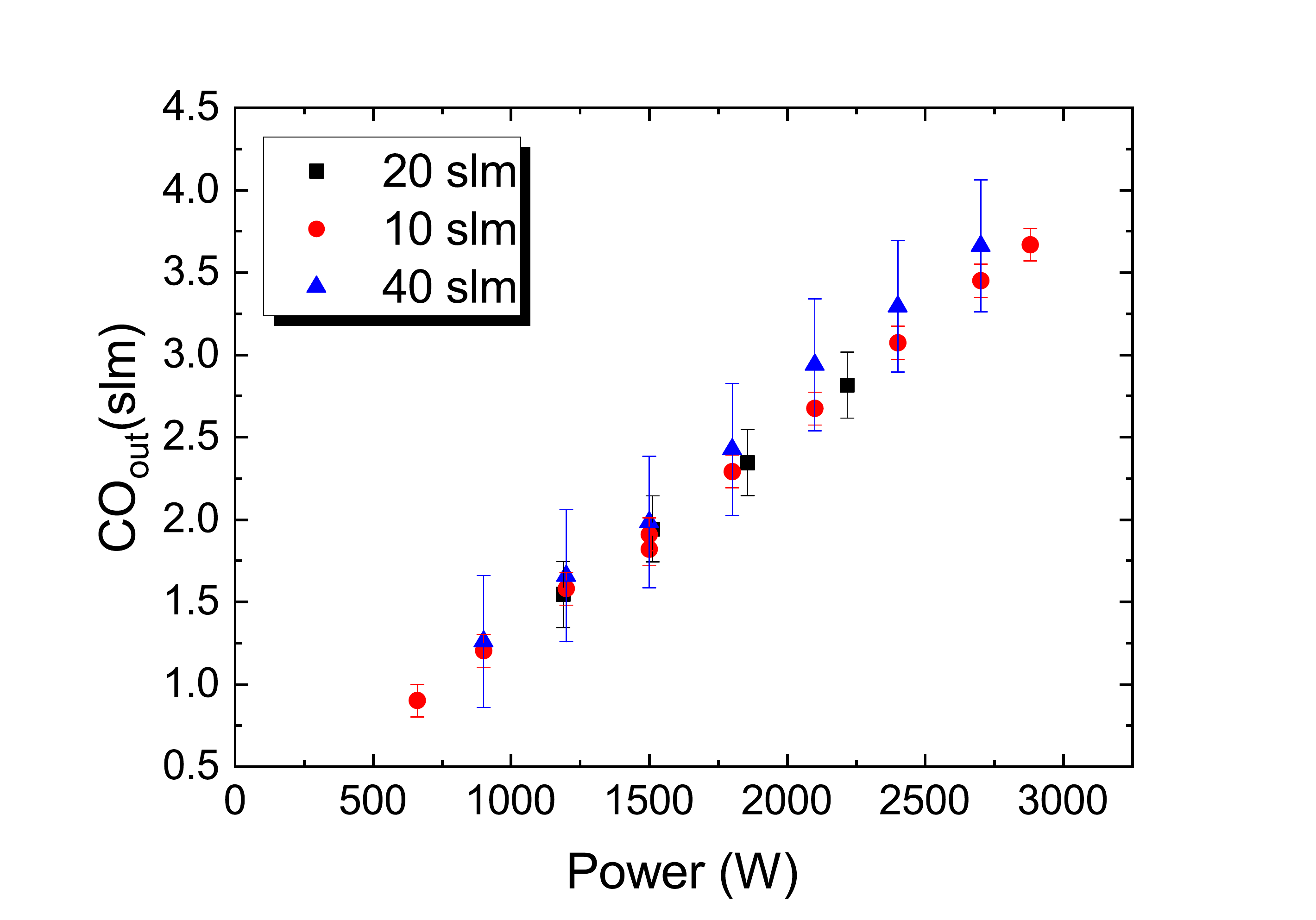}}
\subfloat[][\emph{quasi-atmospheric pressure}]{\includegraphics[width=0.48\textwidth]{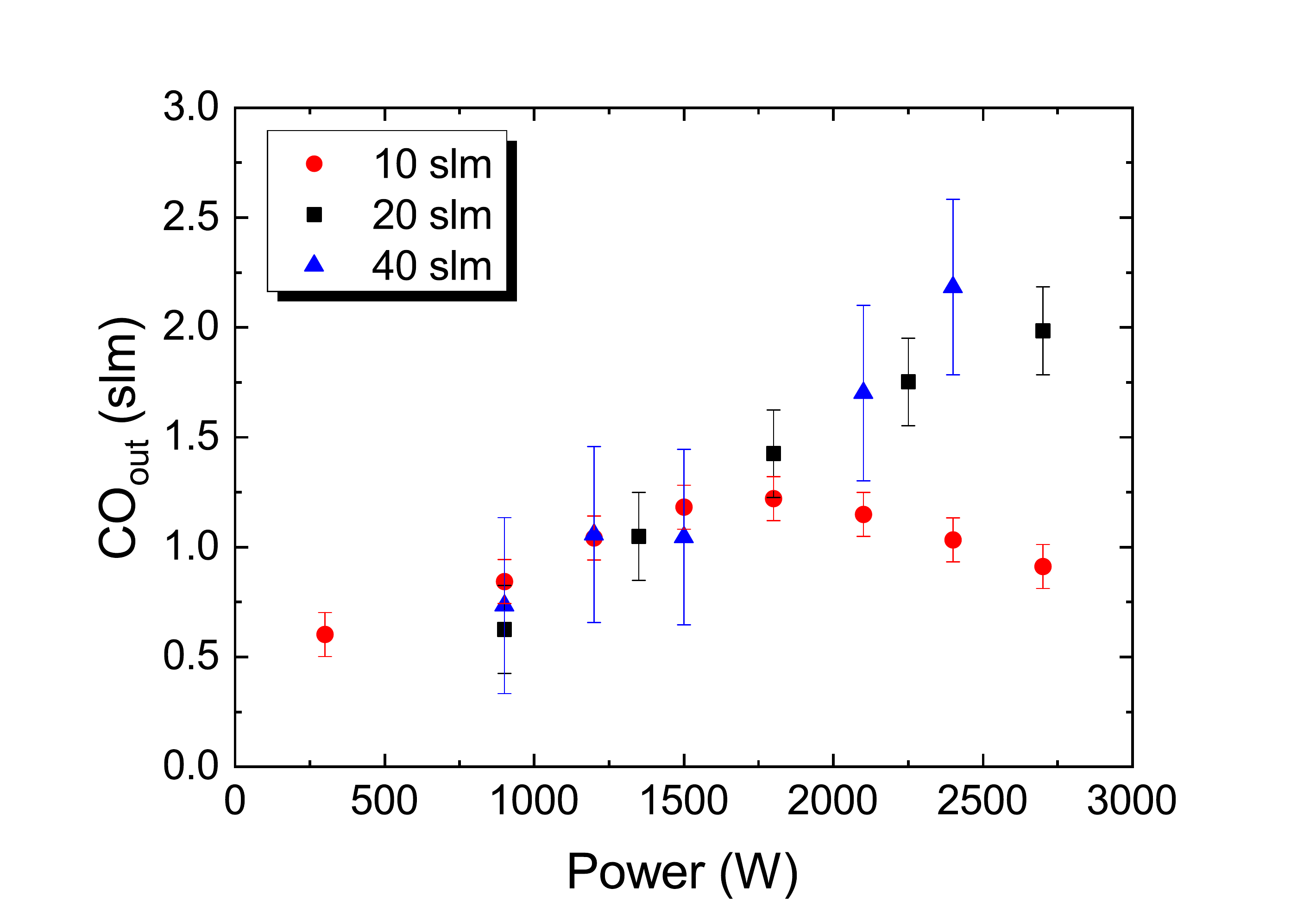}}

\caption{Figure \textit{(a)} and \textit{(b)} show the CO out flow as function of the plasma surface at 200 and 900 mbar, respectively.}
\label{fig:Surf-CO}
\end{figure}

Figures \ref{fig:Surf-CO} \textit{(a)} and \textit{(b)} show the CO$_{out}$ flow at 200 mbar as function of the power coupled into the plasma. Similar trend at 900 mbar is found for the 20 and 40 L/min conditions (figure \ref{fig:Surf-CO}) while for an input gas flow of 10 L/min a reduction in CO$_{out}$ flow is measured. Such reduction of conversion with pressure at the lowest flow appears at 500~mbar and quasi-atmospheric pressure as seen in figure \ref{fig:MS}. The output flow of CO appears to be only a function of the MW power (that scales linearly with the plasma surface, see figures \ref{fig:pscandiameter} and \ref{fig:pscanaxial}) and weakly influenced by the input CO$_2$ gas flow. The turbulent mixing that drives the \COO gas into the plasma is not depending on the magnitude of the \COO flow, thus it is driven by the gas temperature gradient. The approximated temperature gradient $\frac{T_{plasma edge}-T_{wall}}{R_{tube}-R_{plasma}}$ is independent from the gas flow (see section \ref{sec:3.2} and \ref{sec:3.3}). This hypothesis is consistent with the observations carried out in sections \ref{sec:3.2} and \ref{sec:3.3}: neither the plasma size nor the temperature are affected by the \COO input flow.  As a result the influx of cold gas into the plasma region is the same, thus the total CO out flow is constant at given pressure and power. On the other hand a power variation changes the plasma dimensions (see section \ref{sec:3.3}) and consequently the \COO flux into the plasma. 

 
The increase of pressure leads to a significant increase in length of the plasma (see figure \ref{fig:pscanaxial}). 
The increase of \COO influx (due to the longer plasma) is not sufficient to compensate for the higher recombination of \COO into CO (reaction \ref{eq:CORec}) that are expected to increase at higher pressure. At low gas flow and high pressure, where a decrease of conversion is observed at increasing powers (see figure \ref{fig:MS} \fa and \ref{fig:Surf-CO} \fb), the decrease in energy efficiency is then (probably) related to the gas flow dynamics, both in terms of mixing but also heating of the effluent (due to a combination of longer plasma and lower amount of cold \COO gas reservoir) that causes higher recombination of CO into \COO. The present considerations can then only serve as a basis for a detailed modelling of the plasma and gas flow interactions, while focusing on the understanding the impact of the plasma size onto the flow dynamics. The investigation of the CO out flow allows to conclude that the amount of CO produced in the setup used in this work is determined by the pressure and the power coupled into the plasma. 

\subsection{Non-equilibrium considerations}
 An analysis of the \CCd molecule rovibrational distribution functions, performed by Carbone et al. in the same setup \cite{C2paper}, showed that rotational and vibrational states are in equilibrium with the gas temperature. The measured temperatures allow describing the conversion rates of \COO into CO satisfactorily by invoking  a combination of thermal dissociation and flow dynamics. Pietanza et al \cite{pietanza2016influence} performed calculations of the electron energy distribution function (EEDF) in CO$_2$ microwave plasmas while taking into account superelastic processes. They showed that at reduced electric fields typical for microwave discharges that electron are in non-equilibrium with vibrational distribution functions of CO$_2$. Recent plasma impedance simulations carried out by Groen et al. \cite{Groen2019} predict, that in the expanded regime (i.e. at p$<$125 mbar), the electron temperature ranges between 2 and 3 eV. On the other hand, in the contracted regime the electron temperature may vary between 0.5 and 1 eV. Such investigations suggest that the plasma generated in the plasma torch is not in thermal equilibrium. 
The electron impact processes are typically (although those are not the only processes) responsible for the formation of both atomic and molecular electronically excited species. Plasma ground states species calculated from the thermal equilibrium can be used as input for constructing a Boltznamm plot. The density of these species can be correlated to the electron temperature and the EEDF. In addition to rotational and vibrational states densities, the density of electronic states density can be determined by absolute calibrated optical emission spectroscopy.

\begin{figure}[H]
\centering
\subfloat[][\emph{Excitation temperature}]{\includegraphics[width=0.48\textwidth]{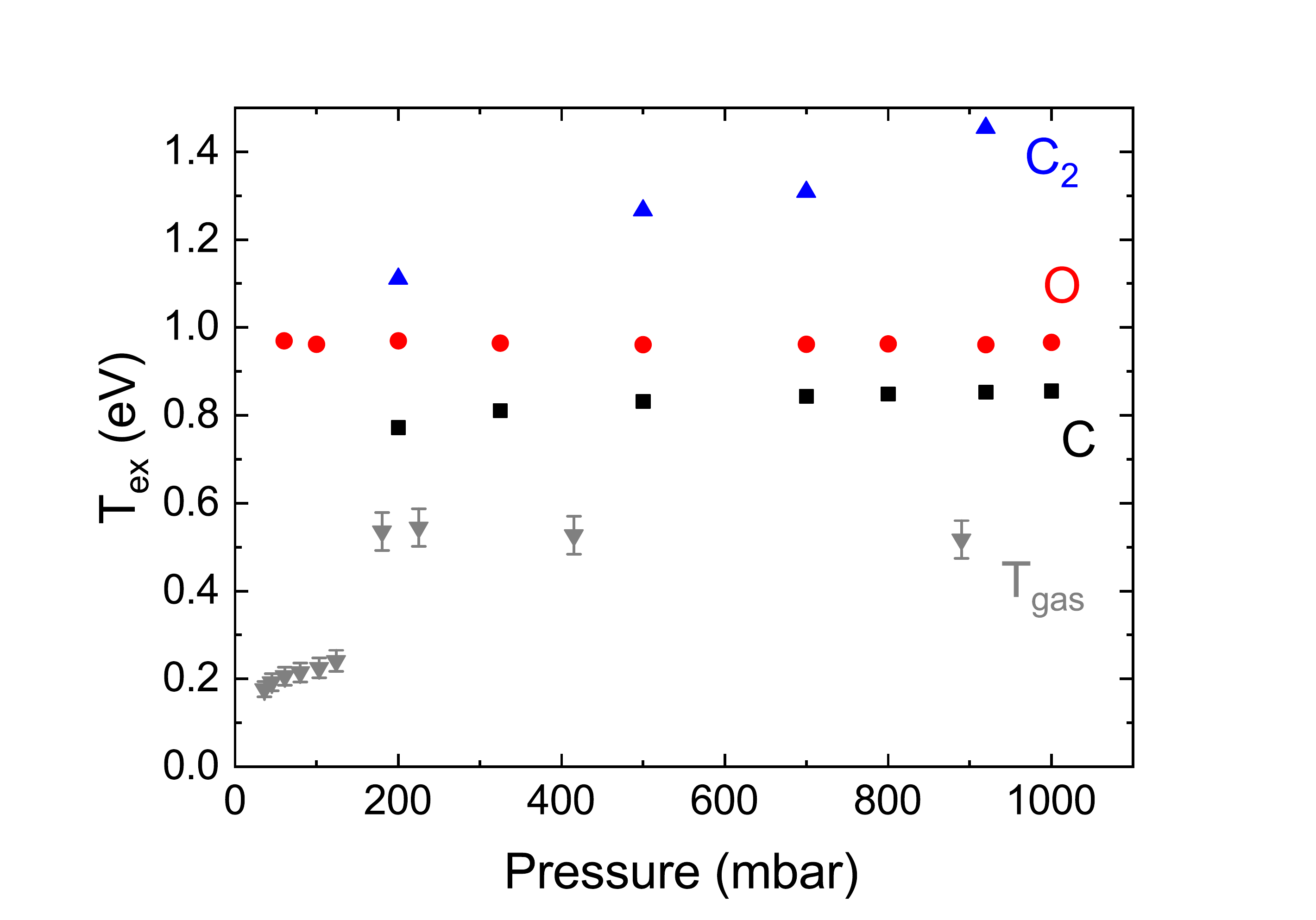}}
\subfloat[][\emph{Typical Boltzmann plot}]{\includegraphics[width=0.48\textwidth]{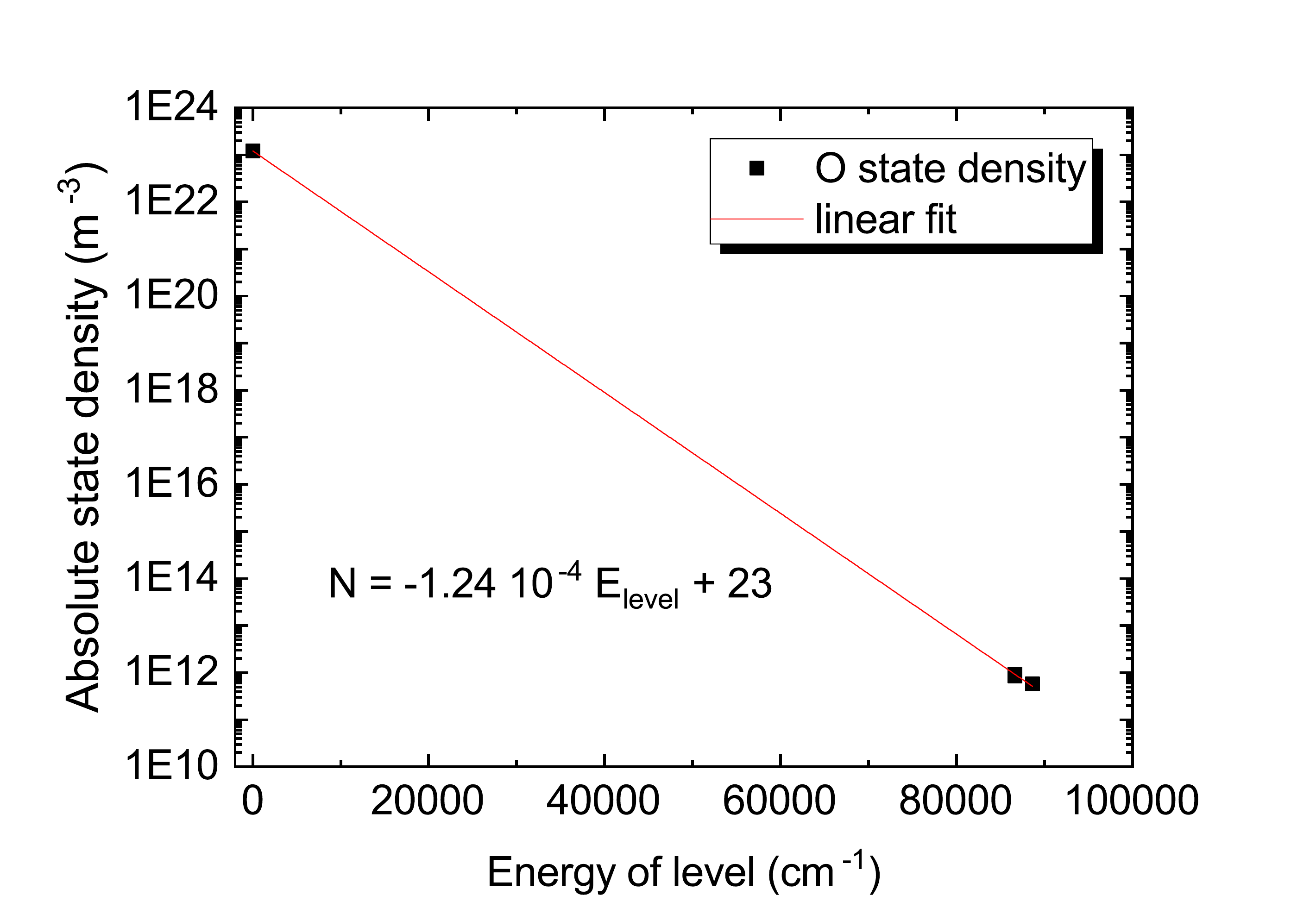}}
\caption{Figure \fa $\:$ shows the excitation temperature of the atomic carbon (black dots), atomic oxygen (red dots), C$_2$ molecule (blue triangle) and gas temperature (gray diamonds). Figure \fb $\:$ the typical Botzmann plot. The one shown here has been obtained at 10~slm, atmospheric pressure and 900~W.} 
\label{fig:Tex}
\end{figure}

Knowing the density of the ground state, it is then possible to define an excitation temperature for that state that is related to the electron temperature. For probing the EEDF, the electronic states need to be predominantly produced by electron impact processes. When the latter condition is not fulfilled (which is likely the case for \CCd state) no information of the EEDF can be obtained. Although this temperature is usually not the electron temperature, its value usually gives an indication of the latter with the electron temperature usually higher than the excitation temperature for ionizing plasma \cite{Joost1990}. Assuming that the chemical composition of the plasma can be described by the gas temperature and chemical equilibrium, the density of ground state can be calculated and a Boltzmann plot constructed. Figure \ref{fig:Tex} \fb $\:$ shows the typical Boltzmann plot that can be obtained combining thermal equilibrium calculation and absolute calibrated optical emission spectroscopy. The shown Boltzmann plot has been obtained at quasi-atmospheric pressure, 10~slm and 900~W. The fitting parameters are shown in figure \ref{fig:Tex} \fb. Figure \ref{fig:Tex} \fa $\:$ shows the excitation temperature for measured electronic states of O and C atoms and the \CCd molecular state as function of the pressure. The transition considered are the C (3s $^1$P $\rightarrow$ 2p$^1$S), the C$_2$ (\Swan)  and the O(3p $^5$P$_{1,2,3}$ $\rightarrow$ 2s$^5$S$_2$), O(3p $^3$P$_{0,1,2}$ $\rightarrow$ 3s$^3$S$_1$). The data points obtained at pressure below 150~mbar are obtained in expanded regime, where no emission from C and C$_2$ can be measured. Excitation temperatures between 0.8 and 1.4 eV are obtained and they are significantly higher than the gas temperature.  

The temperature obtained from O and C excited states are quite close to each other and appear to be insensitive to pressure. The differences between the excitation temperatures of different species can be explained by considering selective population (or depopulation) processes of the excited states. The EEDF does not follow a Boltzmann distribution \cite{pietanza2016influence} and collisional quenching rates are different depending on the species. 
Also, in the case of the \CCd state, this state is the initial source term for the C$_2$ molecule via the reaction \cite{C2paper}

$$C + C + M \rightarrow  C_2 (d^3\Pi_g) + M$$

Note that this source term for \CCd is pressure dependent as it is a three body process. The density of the \CCd increase quadratically with pressure as discussed in Carbone et al. \cite{C2paper} and this is a probable indication of its formation mechanism. Therefore no useful information on the EEDF can be obtained from the excitation temperature of the C$_2$ molecule. If the excitation temperature obtained from the C$_2$ molecule is discarded an electron temperature between 0.8 and 1~eV can then be deduced in the present conditions from the other species. Such electron temperatures are in relatively good agreement with calculations of Groen et al. for the contracted regime \cite{Groen2019}. It can be concluded  that the torch is not fully in local thermal equilibrium and that the electron temperature is higher than the gas temperature, even at atmospheric pressure. 

\section{Conclusions}

A 2.45 GHz microwave \COO plasma torch is characterized by means of OES, iCCD imaging and mass spectrometry in the pressure range between 60 mbar and quasi-atmospheric pressure ($>$900~mbar). At pressures below \ca 120 mbar the plasma is observed to fill the tube while at pressures above, the plasma contracts into a filament, occupying less than 10 \% of the quartz tube cross section. The pressure at which the contraction occurs varies between 110~mbar at 2400~W of input power and 170~mbar at 750~W of input power. The gas temperature increases from to 2400~K up to 3000~K between 60~mbar and \ca 110~mbar. At the contraction point, the gas temperature abruptly increases to 6000~K and remains constant at this value up to quasi-atmospheric pressure. For the gas temperatures reported in this study, the observed \COO conversion rates into CO can be explained solely while invoking thermal processes. Moreover the appearance of the C$_2$ Swan band is correlated with the formation of carbon atoms from CO dissociation that takes place thermally only at temperatures above 5000~K. The highest energy efficiency (\ca 30~\%) is measured at \ca 120~mbar concurrently with the plasma contraction. 


In the contracted regime, the observed \COO conversion rates appear to be driven by the mixing between the cold gas and the hot gas. Indeed, the measured plasma cross section is not sufficient to explain alone the outflow of CO. At 200 mbar, it is found that the CO outflow is independent of the inlet CO$_2$ gas in-flow and that CO outflow scales with input power the latter is correlated to the surface of interactions between the plasma and the cold swirl gas flow and it increases with the power input. Near atmospheric pressure and at low gas flow rates, the conversion rate decreases with input power due to a combination of factors, i.e. swirl and plasma lengths but also gas cooling rates that may not be high enough as well as microwave leaks that can reduce the power coupled to the plasma. The excitation temperatures calculated from the population of the excited species of C, O and C$_2$ indicate that the plasma is not in full local thermal equilibrium and that the electrons have higher temperatures (0.8-1 eV or higher) that the surrounding gas. The measured conversion rates can be explained by considering only thermal dissociation and flow dynamics. Consequently non-equilibrium vibrational ladder climbing leading to \COO dissociation is not needed to explain the present results.

The gas flow dynamics play an important role in the investigated microwave setup, to improve the understanding of the \COO conversion in the present configuration detailed gas flow simulations are necessary. The present experimental results could allow a detailed benchmark of a plasma/gas flow self-consistent calculations in the future. 



\printbibliography

@article{osti_5529228,
title = {Liquid-phase methanol. Final report},
author = {Sherwin, M. and Blum, D.},
abstractNote = {This report is the final report in the RP317 series. This project dealt with the development of the Liquid-Phase Methanol Reactor. The concept of this reactor involves carrying out the synthesis of methanol in an ebullated bed of catalyst in the presence of an inert liquid heat carrier. Work was carried out on laboratory, continuous bench-scale, and a process development unit. Previous reports on this project included AF202 and AF693.},
doi = {},
journal = {},
number = {},
volume = {},
place = {United States},
year = {1979},
month = {12}
}

@Article{Tsang1986,
  author    = {W. Tsang and R. F. Hampson},
  title     = {Chemical Kinetic Data Base for Combustion Chemistry. Part I. Methane and Related Compounds},
  journal   = {Journal of Physical and Chemical Reference Data},
  year      = {1986},
  volume    = {15},
  number    = {3},
  pages     = {1087--1279},
  month     = {jul},
  doi       = {10.1063/1.555759},
  publisher = {{AIP} Publishing},
}

@PhdThesis{PhDLeins,
  author = {Martina Leins},
  title  = {Development and spectroscopic investigation of a microwave plasma source for the decomposition of waste gases},
  school = {Fakultät Mathematik und Physik der Universität Stuttgart},
  year   = {2010},
}

@article{MassiveOES,
  author={Jan Vor{\'a}{\v c} and Petr Synek and Lucia Poto{\v c}{\v n}{\'a}kov{\'a} and Jaroslav Hnilica and V{\'i}t Kudrle},
  title={Batch processing of overlapping molecular spectra as a tool for spatio-temporal diagnostics of power modulated microwave plasma jet},
  journal={Plasma Sources Science and Technology},
  volume={26},
  number={2},
  pages={025010},
  url={http://stacks.iop.org/0963-0252/26/i=2/a=025010},
  year={2017}
}

@Article{C2paper,
  author  = {Emile Carbone and Federico D'Isa and Ante Hecimovic and Ursel Fantz},
  title   = {Analysis of the C$_2$ (d$^3\Pi_g$-a$^3\Pi_u$) Swan bands as a thermometric probe in CO$_2$ microwave plasmas},
  journal={arXiv preprint arXiv:1911.13121},
  year={2019}
}

@Article{MSpaper,
  author  = {Ante Hecimovic and Federico D'Isa and Emile Carbone and Ursel Fantz},
  title   = {Gas composition analysis method for a wide pressure range up to atmospheric pressure - CO$_2$ case study},
  journal = {Review of Scientific Instruments},
  year    = {2019},
 note   = {To be submitted}
}

@Article{BROOKE,
  author   = {James S.A. Brooke and Peter F. Bernath and Timothy W. Schmidt and George B. Bacskay},
  title    = {Line strengths and updated molecular constants for the C$_2$ Swan system},
  journal  = {Journal of Quantitative Spectroscopy and Radiative Transfer},
  year     = {2013},
  volume   = {124},
  pages    = {11 - 20},
  issn     = {0022-4073},
  doi      = {http://dx.doi.org/10.1016/j.jqsrt.2013.02.025},
  url      = {http://www.sciencedirect.com/science/article/pii/S0022407313000800},
}

@Article{Bongers2017,
  author   = {Bongers, Waldo and Bouwmeester, Henny and Wolf, Bram and Peeters, Floran and Welzel, Stefan and van den Bekerom, Dirk and den Harder, Niek and Goede, Adelbert and Graswinckel, Martijn and Groen, Pieter Willem and Kopecki, Jochen and Leins, Martina and van Rooij, Gerard and Schulz, Andreas and Walker, Matthias and van de Sanden, Richard},
  title    = {Plasma-driven dissociation of CO$_2$ for fuel synthesis},
  journal  = {Plasma Processes and Polymers},
  year     = {2017},
  volume   = {14},
  number   = {6},
  pages    = {e201600126--n/a},
  issn     = {1612-8869},
  note     = {e201600126},
  doi      = {10.1002/ppap.201600126},
  url      = {http://dx.doi.org/10.1002/ppap.201600126},
}

@Book{Fridman2008,
  title     = {Plasma chemistry},
  publisher = {Cambridge University Press},
  year      = {2008},
  author    = {Fridman, Alexander},
}

@Article{Snoeckx2017,
  author    = {Snoeckx, Ramses and Bogaerts, Annemie},
  title     = {Plasma technology – a novel solution for CO$_2$ conversion?},
  journal   = {Chem. Soc. Rev.},
  year      = {2017},
  volume    = {46},
  pages     = {5805-5863},
  doi       = {10.1039/C6CS00066E},
  issue     = {},
  publisher = {The Royal Society of Chemistry},
  url       = {http://dx.doi.org/10.1039/C6CS00066E},
}

@Article{Britun2018,
  author   = {Nikolay Britun and Tiago Silva and Guoxing Chen and Thomas Godfroid and Joost van der Mullen and Rony Snyders},
  title    = {Plasma-assisted CO$_2$ conversion: optimizing performance via microwave power modulation},
  journal  = {Journal of Physics D: Applied Physics},
  year     = {2018},
  volume   = {51},
  number   = {14},
  pages    = {144002},
  url      = {http://stacks.iop.org/0022-3727/51/i=14/a=144002},
}

@Article{Leins2013,
  author    = {M. Leins and J. Kopecki and S. Gaiser and A. Schulz and M. Walker and U. Schumacher and U. Stroth and T. Hirth},
  title     = {Microwave Plasmas at Atmospheric Pressure},
  journal   = {Contributions to Plasma Physics},
  year      = {2013},
  volume    = {54},
  number    = {1},
  pages     = {14--26},
  doi       = {10.1002/ctpp.201300033},
  publisher = {Wiley},
}

@Article{Belov2018,
  author    = {Igor Belov and Vincent Vermeiren and Sabine Paulussen and Annemie Bogaerts},
  title     = {Carbon dioxide dissociation in a microwave plasma reactor operating in a wide pressure range and different gas inlet configurations},
  journal   = {Journal of {CO}$_2$ Utilization},
  year      = {2018},
  volume    = {24},
  pages     = {386--397},
  doi       = {10.1016/j.jcou.2017.12.009},
  publisher = {Elsevier {BV}},
}

@Article{Mitsingas2016,
  author    = {Constandinos M. Mitsingas and Rajavasanth Rajasegar and Stephen Hammack and Hyungrok Do and Tonghun Lee},
  title     = {High Energy Efficiency Plasma Conversion of CO$_2$ at Atmospheric Pressure Using a Direct-Coupled Microwave Plasma System},
  journal   = {{IEEE} Transactions on Plasma Science},
  year      = {2016},
  volume    = {44},
  number    = {4},
  pages     = {651--656},
  doi       = {10.1109/tps.2016.2531641},
  publisher = {Institute of Electrical and Electronics Engineers ({IEEE})},
}

@Article{Spencer2012,
  author    = {L. F. Spencer and A. D. Gallimore},
  title     = {CO$_2$ dissociation in an atmospheric pressure plasma catalyst system: a study of energy efficiency},
  journal   = {Plasma Sources Science and Technology},
  year      = {2012},
  volume    = {22},
  number    = {1},
  pages     = {015019},
  month     = dec,
  doi       = {10.1088/0963-0252/22/1/015019},
  publisher = {{IOP} Publishing},
}

@Article{Drenik2017,
  author    = {Aleksander Drenik and Daniel Alegre and Sebastijan Brezinsek and Alfonso de Castro and Uron Kruezi and Gerd Meisl and Miran Mozetic and Martin Oberkofler and Matjaz Panjan and Gregor Primc and Matic Resnik and Volker Rohde and Michael Seibt and Francisco L. Tabar{\'{e}}s and Rok Zaplotnik},
  title     = {Detection of ammonia by residual gas analysis in {AUG} and {JET}},
  journal   = {Fusion Engineering and Design},
  year      = {2017},
  volume    = {124},
  pages     = {239--243},
  month     = nov,
  doi       = {10.1016/j.fusengdes.2017.05.037},
  publisher = {Elsevier {BV}},
}

@Article{Groen2019,
  author    = {Petrus Wilhelmus Cornelis Groen and Abraham J. Wolf and Tim W.H. Righart and Richard Van de Sanden and Floran Peeters and Waldo Bongers},
  title     = {Numerical model for the determination of the reduced electric field in a CO$_2$ microwave plasma derived by the principle of impedance matching},
  journal   = {Plasma Sources Science and Technology},
  year      = {2019},
  month     = apr,
  doi       = {10.1088/1361-6595/ab1ca1},
  publisher = {{IOP} Publishing},
}

@Article{Rooij2015,
  author    = {G. J. van Rooij and D. C. M. van den Bekerom and N. den Harder and T. Minea and G. Berden and W. A. Bongers and R. Engeln and M. F. Graswinckel and E. Zoethout and M. C. M. van de Sanden},
  title     = {Taming microwave plasma to beat thermodynamics in CO$_2$ dissociation},
  journal   = {Faraday Discussions},
  year      = {2015},
  volume    = {183},
  pages     = {233-248},
  doi       = {10.1039/c5fd00045a},
  publisher = {Royal Society of Chemistry ({RSC})},
}

@Article{Du2017,
  author   = {Du, Yanjun and Tamura, Keishiro and Moore, Sampson and Peng, Zhimin and Nozaki, Tomohiro and Bruggeman, Peter J.},
  title    = {CO(B1$\Sigma^+\rightarrow A^1\Pi$) Angstrom System for Gas Temperature Measurements in CO$_2$ Containing Plasmas},
  journal  = {Plasma Chemistry and Plasma Processing},
  year     = {2017},
  volume   = {37},
  number   = {1},
  pages    = {29--41},
  month    = jan,
  issn     = {1572-8986},
  day      = {01},
  doi      = {10.1007/s11090-016-9759-5},
  url      = {https://doi.org/10.1007/s11090-016-9759-5},
}

@Article{Babou2008,
  author    = {Yacine Babou and Philippe Rivi{\`{e}}re and Marie-Yvonne Perrin and Anouar Soufiani},
  title     = {Spectroscopic study of microwave plasmas of CO$_2$ and CO$_2$-N$_2$ mixtures at atmospheric pressure},
  journal   = {Plasma Sources Science and Technology},
  year      = {2008},
  volume    = {17},
  number    = {4},
  pages     = {045010},
  month     = aug,
  doi       = {10.1088/0963-0252/17/4/045010},
  publisher = {{IOP} Publishing},
}

@Article{Silva2014,
  author    = {Tiago Silva and Nikolay Britun and Thomas Godfroid and Rony Snyders},
  title     = {Simple method for gas temperature determination in CO$_2$ containing discharges},
  journal   = {Optics Letters},
  year      = {2014},
  volume    = {39},
  number    = {21},
  pages     = {6146},
  month     = oct,
  doi       = {10.1364/ol.39.006146},
  publisher = {The Optical Society},
}

@Article{Bekerom2018,
  author    = {Dirk van den Bekerom and Jose Maria Palomares Linares and Tiny Verreycken and Eddie M Van Veldhuizen and Sander Nijdam and Waldo Bongers and Richard Van de Sanden and Gerard J Van Rooij},
  title     = {The importance of thermal dissociation in CO$_2$ microwave discharges investigated by power pulsing and rotational Raman scattering},
  journal   = {Plasma Sources Science and Technology},
  year      = {2018},
  month     = nov,
  doi       = {10.1088/1361-6595/aaf519},
  publisher = {{IOP} Publishing},
}

@Article{CEA,
  author  = {McBride, Bonnie J and Gordon, Sanford},
  title   = {Computer Program for Calculation of Complex Chemical Equilibrium Compositions and Applications II Users Manual and Program Description. 2; Users Manual and Program Description},
  journal = {NASA Lewis Research Center; Cleveland,OH United States},
  year    = {1996},
}

@Article{Rusanov1984,
  author   = {{Rusanov}, V.~D. and {Fridman}, A.~A.},
  title    = {{The physics of a chemically active plasma}},
  journal  = {Moscow Izdatel Nauka},
  year     = {1984},
}

@Article{LEGASOV1978,
  author   = {{Legasov}, V.~A. and {Zhivotov}, V.~K. and {Krasheninnikov}, E.~G. and {Krotov}, M.~F. and {Patrushev}, B.~I. and {Rusanov}, V.~D. and {Rykunov}, G.~V. and {Spektor}, A.~M. and {Fridman}, A.~A. and {Sholin}, G.~V.},
  title    = {{Nonequilibrium plasma-chemical process of the decomposition of CO$_2$ in HF and UHF discharges}},
  journal  = {Akademiia Nauk SSSR Doklady},
  year     = {1978},
  volume   = {238},
  pages    = {66-69},
  month    = jan,
  note     = {Traslation curtesy of Nicolay Britun},
}

@Article{Butylkin1981,
  author   = {{Butylkin}, I.~P. and {Zhivotov}, V.~K. and {Krasheninnikov}, E.~G. and {Krotov}, M.~F. and {Rusanov}, V.~D. and {Tarasov}, I.~V. and {Fridman}, A.~A.},
  title    = {{Plasma-chemical process of CO$_2$ dissociation in a nonequilibrium microwave discharge}},
  journal  = {Zhurnal Tekhnicheskoi Fiziki},
  year     = {1981},
  volume   = {51},
  pages    = {925-931},
  month    = may,
}

@Article{ASISOV1983,
  author   = {{Azizov}, R.~I. and {Vakar}, A.~K. and {Zhivotov}, V.~K. and {Krotov}, M.~F. and {Zinovev}, O.~A. and {Potapkin}, B.~V. and {Rusanov}, A.~A. and {Rusanov}, V.~D. and {Fridman}, A.~A.},
  title    = {{Nonequilibrium plasmachemical process of CO$_2$ decomposition in a supersonic microwave discharge}},
  journal  = {Akademiia Nauk SSSR Doklady},
  year     = {1983},
  volume   = {271},
  pages    = {94-98},
  month    = aug,
  note     = {Traslation curtesy of Nicolay Britun},
}

@Article{Aresta2013,
  author    = {Michele Aresta and Angela Dibenedetto and Antonella Angelini},
  title     = {The changing paradigm in {CO}$_2$ utilization},
  journal   = {Journal of {CO}$_2$ Utilization},
  year      = {2013},
  volume    = {3-4},
  pages     = {65--73},
  month     = {dec},
  doi       = {10.1016/j.jcou.2013.08.001},
  publisher = {Elsevier {BV}},
}

@Book{ETP2017,
  title     = {Energy Technology Perspectives 2017},
  publisher = {IEA},
  year      = {2017},
  author    = {IEA},
  editor    = {IEA},
  month     = jun,
  isbn      = {978-92-64-27597-3},
}

@Article{Quadrelli2011,
  author    = {Elsje Alessandra Quadrelli and Gabriele Centi and Jean-Luc Duplan and Siglinda Perathoner},
  title     = {Carbon Dioxide Recycling: Emerging Large-Scale Technologies with Industrial Potential},
  journal   = {{ChemSusChem}},
  year      = {2011},
  volume    = {4},
  number    = {9},
  pages     = {1194--1215},
  month     = {sep},
  doi       = {10.1002/cssc.201100473},
  publisher = {Wiley},
}

@Article{Ganesh2014,
  author    = {Ibram Ganesh},
  title     = {Conversion of carbon dioxide into methanol {\textendash} a potential liquid fuel: Fundamental challenges and opportunities (a review)},
  journal   = {Renewable and Sustainable Energy Reviews},
  year      = {2014},
  volume    = {31},
  pages     = {221--257},
  month     = {mar},
  doi       = {10.1016/j.rser.2013.11.045},
  publisher = {Elsevier {BV}},
}

@InCollection{schlogl2013solar,
  author    = {Schl{\"o}gl, Robert},
  title     = {The solar refinery},
  booktitle = {Chemical Energy Storage},
  publisher = {De Gruyter},
  year      = {2013},
  pages     = {1--34},
}

@InProceedings{FloranISPC24,
  author    = {Floran Peeters and Huub Hendrickx and Alex van de Steeg and Tim Righart and Bram Wolf and Gerard van Rooij and Waldo Bongers and Richard van de Sanden},
  title     = {Chemiluminescence as a diagnostic tool in CO$_2$ microwave plasma},
  booktitle = {ISPC 24 Naples},
  year      = {2019},
}

@article{showstack2013carbon,
  title={Carbon dioxide tops 400 ppm at Mauna Loa, Hawaii},
  author={Showstack, Randy},
  journal={Eos, Transactions American Geophysical Union},
  volume={94},
  number={21},
  pages={192--192},
  year={2013},
  publisher={Wiley Online Library}
}

@article{monastersky2013global,
  title={Global carbon dioxide levels near worrisome milestone},
  author={Monastersky, Richard},
  journal={Nature News},
  volume={497},
  number={7447},
  pages={13},
  year={2013}
}

@article{Bogaerts2018ACSEL,
author = {Bogaerts, Annemie and Neyts, Erik C.},
title = {Plasma Technology: An Emerging Technology for Energy Storage},
journal = {ACS Energy Letters},
volume = {3},
number = {4},
pages = {1013-1027},
year = {2018},
doi = {10.1021/acsenergylett.8b00184},

URL = { 
        https://doi.org/10.1021/acsenergylett.8b00184
    
},
eprint = { 
        https://doi.org/10.1021/acsenergylett.8b00184
    
}

}

@article{azizov1983vakar,
  title={{The nonequilibrium plasmachemical process of decomposition of CO$_2$ in a supersonic SHF discharge}},
  author={Azizov, R I and Vakar, A K and Zhivotov, V K and Krotov, M F and Zinov'ev, O A and Potapkin, BV and Rusanov, A A and Rusanov, V D and Fridman, A A},
  booktitle={Soviet Physics Doklady},
  volume={28},
  pages={567},
  year={1983}
}

@article{pietanza2016influence,
  title={{Influence of electron molecule resonant vibrational collisions over the symmetric mode and direct excitation-dissociation cross sections of CO$_2$ on the electron energy distribution function and dissociation mechanisms in cold pure CO$_2$ plasmas}},
  author={Pietanza, LD and Colonna, G and Laporta, V and Celiberto, R and D’Ammando, G and Laricchiuta, A and Capitelli, M},
  journal={The Journal of Physical Chemistry A},
  volume={120},
  number={17},
  pages={2614--2628},
  year={2016},
  publisher={ACS Publications}
}

@article{Joost1990,
title = "Excitation equilibria in plasmas; a classification",
journal = "Physics Reports",
volume = "191",
number = "2",
pages = "109 - 220",
year = "1990",
issn = "0370-1573",
doi = "https://doi.org/10.1016/0370-1573(90)90152-R",
url = "http://www.sciencedirect.com/science/article/pii/037015739090152R",
author = "J.A.M. van der Mullen"
}

@article{Moisan2018,
	doi = {10.1088/1361-6595/aac528},
	year = {2018},
	publisher = {{IOP} Publishing},
	volume = {27},
	number = {7},
	pages = {073001},
	author = {Michel Moisan and Helena Nowakowska},
	title = {Contribution of surface-wave ({SW}) sustained plasma columns to the modeling of {RF} and microwave discharges with new insight into some of their features. A survey of other types of {SW} discharges},
	journal = {Plasma Sources Science and Technology}
}

@Article{Schulz2012,
  author    = {A. Schulz and P. Büchele and E. Ramisch and O. Janzen and F. Jimenez and C. Kamm and J. Kopecki and M. Leins and S. Merli and H. Petto and F.R. Mendez and J. Schneider and U. Schumacher and M. Walker and U. Stroth},
  title     = {Scalable Microwave Plasma Sources From Low to Atmospheric Pressure},
  journal   = {Contributions to Plasma Physics},
  year      = {2012},
  volume    = {52},
  number    = {7},
  pages     = {607--614},
  month     = {aug},
  doi       = {10.1002/ctpp.201210057},
  publisher = {Wiley},
}
\clearpage

\end{document}